\documentclass[fleqn,usenatbib]{mnras}

\usepackage{newtxtext,newtxmath}

\usepackage[T1]{fontenc}
\usepackage{ae,aecompl}

\usepackage{graphicx}	
\usepackage{amsmath}	
\usepackage{amssymb}	
\usepackage{xcolor}

\newcommand{\hii}{\ion{H}{II}}
\newcommand{\SII}{[\ion{S}{II}]}  
\newcommand{\CII}{[\ion{C}{II}]}  
\newcommand{\NII}{[\ion{N}{II}]}  
\newcommand{\AV}{$A_{\rm V}$} 
\newcommand{\Tdust}{$T_{\rm dust}$} 
\newcommand{\Halpha}{H$\alpha$} 
\newcommand{\Hbeta}{H$\beta$} 
\newcommand{\nelec}{$n_{\rm e}$}
\newcommand{\Telec}{$T_{\rm e}$}
\newcommand{\kms}{km s$^{-1}$}
\newcommand{\mkm}{$\mu$m}

\title[3D structure of S235]
{3D structure of the H\,{\Large \textbf{II}} region Sh2-235 from tunable-filter optical observations}

\author[M. S. Kirsanova et al.]{
M. S. Kirsanova,$^{1,2}$\thanks{E-mail: kirsanova@inasan.ru}
P. A. Boley$^{2,3}$,
A. V. Moiseev$^{4}$,
D. S. Wiebe$^{1}$,
R. I. Uklein$^{4}$
\\
$^{1}$Institute of Astronomy, Russian Academy of Sciences, 119017, 48 Pyatnitskaya Str., Moscow, Russia\\
$^{2}$Moscow Institute of Physics and Technology, 141701, 9 Institutskiy per., Dolgoprudny, Moscow Region, Russia\\
$^{3}$Institute of Natural Sciences and Mathematics, Ural Federal University, 19 Mira Str., 620075 Ekaterinburg, Russia\\
$^{4}$Special Astrophysical Observatory, Russian Academy of Sciences, Nizhny Arkhyz 369167, Russia\\
}

\date{Accepted 2020 July 04. Received 2020 July 04; in original form 2019 November 11}

\pubyear{2019}

\begin{document}
\label{firstpage}
\pagerange{\pageref{firstpage}--\pageref{lastpage}}
\maketitle

\begin{abstract}
We present observations of the \Halpha, \Hbeta, \SII~$\lambda\lambda6716, 6731$ and \NII~$\lambda6583$ emission lines in the galactic \hii{} region Sh2-235 with the Mapper of Narrow Galaxy Lines (MaNGaL), a tunable filter at the 1-m telescope of the Special Astrophysical Observatory of the Russian Academy of Sciences. We show that the \hii{} region is obscured by neutral material with $A_{\rm V} \approx 2-4$~mag. The area with the highest \AV{} is situated to the south-west from the ionizing star and coincides with a maximum detected electron density of $\gtrsim300$~cm$^{-3}$. The combination of these results with archive {\it AKARI} far-infrared data allows us to estimate the contribution of the front and rear walls to the total column density of neutral material in S235, and explain the three-dimensional structure of the region. The \hii{} region consist of a denser, more compact portion deeply embedded in the neutral medium and the less dense and obscured gas. The front and rear walls of the \hii{} region are inhomogeneous, with the material in the rear wall having a higher column density. We find a two-sided photodissociation region in the dense clump S235~East~1, illuminated by a UV field with $G_0=50-70$ and 200 Habing units in the western and eastern parts, respectively.
\end{abstract}




\section{Introduction}

The \hii{} region Sh2-235 \citep[S235 hereafter,][]{Sharpless59} is the brightest \hii{} region among four close emission nebulae (the others being Sh2-231, Sh2-232 and Sh2-233) located in the giant molecular cloud G174+2.5 \citep{Ladeyschikov16} in the direction of the Aur~OB1 association \citep{Straizys10} in the Perseus Spiral Arm.  S235 and its surroundings are a region of active star formation \citep[see e.g. recent studies by][]{Chavarria14,Bieging16}. It contains at least five young stellar clusters projected on the border of the \hii{} region, as was initially shown by \citet{Kirsanova08}, and later studied by \citet{Camargo11,Dewangan11,Kirsanova14}. The surroundings of S235, namely the photodissociation region (PDR), molecular envelope, and young stellar objects embedded in the envelope, have been extensively observed at infrared \citep[][]{Allen05, Klein05, Anderson19} and radio wavelengths \citep[][]{Israel1978, 1978ApJ...224..437S,Evans81a, Lafon83, Heyer96, Kirsanova14, Bieging16, Ladeyschikov16, Burns19}, although the \hii{} region itself has not been extensively studied. 

Using radio recombination line (RRL) emission, \citet{Quireza06b,Quireza06a} determined that S235 (their source G173.60+2.80) has a nearly circular shape with an angular diameter of about 5\arcmin, an electron density of \nelec=81.6~cm$^{-3}$ and an electron temperature of $T_{\rm e} = 8940 \pm 170$~K. \citet{Anderson19}, using a combination of carbon and hydrogen RRLs, \CII{} emission at 158~\mkm{} and CO emission determined that the \hii{} region lies on the near side of the associated molecular cloud and is expanding in the direction of the observer. The distance to the ionizing star of S235, BD+35$^\circ$1201, is $1.65\pm0.1$~kpc, based on the Gaia DR2 parallax measurements \citep{Gaia16,Gaia18} and the distance calibration of \citet{BailerJones18}. The \hii{} region is excited by a late O- \citep{Georgelin73} or early B-type \citep{Hunter90} star. \citet{Lafon83} inspected an H$\alpha$ image of the nebula, and found that it consists of a bright northern, and more diffuse southern, part. They also reported a north-south radial velocity gradient of the ionized gas, although this was not confirmed by recent observations of RRLs by \citet{Anderson19}. \citet{Straizys10} determined the interstellar extinction towards the ionizing star to be $A_{\rm V} = 3.7$~mag. \citet{Esteban18} measured the physical parameters of the nebula with optical slit spectroscopy, and found \nelec{} = 120~cm$^{-3}$ from the \SII{}~$\lambda\lambda$6716, 6731 lines and $T_{\rm e}$ in the range from 7100 to 11900~K, depending on the line ratio used (\NII, \SII, [\ion{O}{II}] or [\ion{O}{III}] lines). These optical observations were confined to a single slit position with a length of 30\arcsec{}, and therefore do not give a full overview of the properties of the ionized gas in this extended \hii{} region.

The aim of the present work is to study the properties and spatial distribution of the ionized gas in S235 using narrow-band images in optical forbidden and Balmer emission lines, obtained with a new tunable-filter photometer at the 1-m Zeiss-1000 telescope of the Special Astrophysical Observatory of the Russian Academy of Sciences (SAO RAS), supplemented by additional imaging and spectroscopic data from space- and ground-based telescopes (\textit{AKARI} infrared imaging satellite, SAO RAS 6-m telescope, and previously published CO maps).

\section{Observations and data reduction}

\subsection{MaNGaL imaging}

We performed observations using the Zeiss-1000 telescope of SAO~RAS with the Mapper of Narrow Galaxy Lines (MaNGaL). MaNGaL is a new tunable-filter photometer developed at SAO RAS, based on a piezoelectric scanning Fabry-Perot interferometer (FPI) with low interference order ($n\approx20$ in the \Halpha{} line). The instrument was described by \citet{mangal} and at the SAO RAS web-page\footnote{\url{https://www.sao.ru/Doc-en/Events/2017/Moiseev/moiseev_eng.html}}. First results using the instrument were presented by \citet{Keel2019}. 
The width of the instrumental profile in the spectral range used was $FWHM=14\pm1$\AA, and the central wavelength ($CWL$) of the peak of FPI transmission was precisely tuned to the desired wavelength (better than 0.5~\AA) at the centre of the field of view. For the observations in January 2018 we used an Andor iKON-M 934 CCD Camera, which provides a field of view of 8.7\arcmin{} and a plate scale 0.51\arcsec{}/px. 

Unlike the `classical' optical layout having a tunable filter in the collimated beam \citep[e.g.][]{Jones2002}, MaNGaL is an afocal reducer with the FPI in the convergent beam \citep{Courtes1960}. This arrangement provides a significantly larger field for the central monochromatic region (`Jacquinot spot'), which is crucial for studying extended targets. In our case, $CWL$ varies with a range smaller than the filter $\pm0.5FWHM$ across the whole field of view.  As a result, variations of the FPI transmission of the nebular emission lines in S235 should not be very significant, because their line-of-sight velocity variations are smaller than 100~\kms (Sec.~\ref{sec:spectra}). Nevertheless, we applied a correction for the effects of $CWL$ variations (see below).   

We show the observation log of the MaNGaL observations in Table~\ref{tab:Z1000_obs}, where the seeing is given for the resulting combination of all frames. The most deep and detailed data in the \Halpha{} and \SII{} lines were obtained in January, 2018, in the \textit{`single images FPI mode'}, where the $CWL$ is tuned first to the emission line (taking into account the systemic velocity of the target and heliocentric correction), and then to the neighbouring continuum (shifted by $30-50$\AA). This cycle was repeated, which averages the contribution of atmospheric seeing and variations in atmospheric transparency. The FPI transmission peaks from neighbouring interference orders were fully blocked using medium-width filters with a bandwidth of $\sim250$~\AA. Different blocking filters were used for \Halpha{} and \SII{} spectral ranges. In the case of the \SII{} doublet, the continuum was observed both redward and blueward of the emission lines; for \Halpha{}, the continuum was observed only on the blue side.

In order to understand how these single images were affected by the mismatch between the FPI peak $CWL$ and the emission line centre, immediately after taking deep images we performed observations in the \textit{`scanning FPI mode'}; i.e. we quickly scanned the wavelength regions around the \Halpha+\NII{} and \SII{} emission lines: 12 subsequent frames with $CWL$ increments of 7.5~\AA{} were obtained for each spectral interval with relatively short (60~s) exposures, to minimize the effects of atmospheric variations. The CCD was operated with $4\times4$ binning in order to obtain a signal-to-noise (S/N) ratio comparable to that of the single images. 

In September 2018, we used an Andor Neo 5.5 sCMOS camera, because the primary CCD detector was under repair. With $2\times2$ binning the Neo sCMOS provides the same sampling ($0.5$\arcsec/px) and field of view, but with lower quantum efficiency and significantly higher noise compared to the iKON CCD.  For these observations MaNGaL was operated in the \textit{`direct images mode'} like a standard photometer, without the FPI. Images were exposed in medium-band filters with $FWHM\sim100$~\AA{} centred on the \Hbeta{} emission line (filter $CWL\approx$4880~\AA) and continuum near ($CWL\approx$5150~\AA).  

The data reduction was performed using a custom software package running in the IDL environment\footnote{\url{https://www.harrisgeospatial.com/Software-Technology/IDL}}, which includes bias (for CCD) and dark current (for sCMOS) subtraction,  flat-field correction, and cosmic ray removal by combining individual short exposures at the same wavelength.  Continuum emission was removed from the images in the lines by normalising the continuum to minimise the flux residuals in the background and foreground stars. Per-pixel uncertainties were derived based on photon statistics in the line and continuum images. The typical value of the background RMS of the net emission line images was $(3-6)\times10^{-17}$~ergs~cm$^{-2}$sec$^{-1}$arcsec$^{-2}$. The quality of the continuum subtraction in the single images obtained with the FPI mode was significantly better than in those obtained in the `classical' filter direct images in \Hbeta, due to the fact that the continuum was observed at very close wavelengths and with a much narrower filter width. 

In order to calibrate the images to an absolute intensity scale, each night we observed spectrophotometric standard stars in the corresponding observing mode immediately before or after S235, and at a similar airmass. In January, 2018 we observed the standard star G191B2B at a zenith distance of $z=25$\degr{}, which corresponds to an airmass of 1.10, whereas S235 was observed in the airmass range 1.04--1.19. In September, 2018 the nebula was observed in the larger airmass range 1.05--1.66, while the standard stars Hz2 and Hz4 were exposed at zenith distances corresponding to airmasses of 1.04, 1.16 and 1.74. All nights were photometric. The equations from \citet{Jones2002} were used for calibration of our tunable-filter data. The SAO RAS extinction curve \citep{Kartasheva1978} was used for atmospheric extinction correction. The difference between the S235 \Hbeta{} fluxes estimated in different nights in September, 2018 was about 4\%. This value can be considered as the accuracy of our flux calibration.

The astrometric calibration was performed using astrometry.net \citep{Lang2010} and index files compiled from the Gaia DR2 catalog \citep{Gaia18}. Using 56 Gaia stars in the field brighter than $G=17$~mag, we find a position RMS of the astrometric solution of 0.12\arcsec{}.

For the wavelength calibration and analysis of the images obtained in the scanning mode, we modified the IDL-based software for scanning FPI data reduction of \citet{MoiseevEgorov2008}. The images were merged into data cubes containing a 12-channel spectrum at each pixel. These low-resolution spectra were fitted at each position in the field by a combination of Gaussians, corresponding to the \Halpha, \NII$\lambda\lambda6548,6583$ and \SII$\lambda\lambda6716,6731$ emission lines. The wavelength difference between lines, $FWHM$  and \NII{} doublet ratio (1:3) were fixed, while the radial velocity and line amplitudes were left as free parameters.  After this procedure we have two types of $4\times4$ binning maps in the \Halpha{} and \SII{} lines:  (1) the fitting results ($F_{\rm fit}$), free from $CWL$-emission peak mismatch; (2) the data cube channels ($F_{\rm image}$) observed with the same parameters of the FPI as the corresponding \textit{`single images'}. The ratio of $F_{\rm fit}/F_{\rm image}$ after smoothing and interpolation was used to correct the emission line images in the original $1\times1$ binning resolution.  In most areas of the field the correction is negligible ($F_{\rm fit}/F_{\rm image}=0.95-1.1$), and it exceeded a factor of 1.2 only in the northwest edge of the field, where a small emission filament is located. The uncertainty of this ratio estimated as an RMS value on the non-smoothed maps was 0.03--0.05 in the main field, and about 0.1--0.2 at the northwest edge of the nebula.

This type of correction only affects the absolute flux distribution and \Hbeta$/$\Halpha{} ratio, and does not affect the line ratio maps taken in the \textit{`single images'} mode: \SII$/$\Halpha, \SII$\lambda6716/$\SII$\lambda6731$. As a side result, the data cube fitting also provided an emission map of the \NII$\lambda6583$ line.    

\begin{table}
	\centering
	\caption{MaNGaL observations at the SAO RAS Zeiss-1000 telescope}
	\label{tab:Z1000_obs}
	\begin{tabular}{llcr} 
		\hline
		Date  & Sp. range& Exposure, s &  seeing, $''$\\
		\hline
	    \multicolumn{4}{c}{\it Mode: single images with FPI}\\
    2018 Jan 26& \Halpha          & 1200&  1.9 \\
              & \Halpha{} continuum& 1200&  1.9 \\
              &  \SII$\lambda6716$& 2700 & 1.9\\
              & \SII$\lambda6731$ & 5400 & 1.9\\
              & \SII{} continuum 1& 2700   & 1.9\\
              & \SII{} continuum 2& 2700   & 1.9\\
        	\hline      
	    \multicolumn{4}{c}{\it Mode: scanning with FPI}\\
	2018 Jan 26& \Halpha+\NII  & $12\times60$&  1.9 \\
		2018 Jan 26& \SII   & $12\times60$&  1.9 \\
		\hline
	    \multicolumn{4}{c}{\it Mode: direct images}\\
	2018 Sep 18& \Hbeta                      & 4800&  2.5 \\
	         & \Hbeta{} continuum           & 5400&  2.5 \\	    
 	2018 Sep 19& \Hbeta                      & 5100&  1.9 \\
	          & \Hbeta{} continuum          & 3900&  1.9 \\
 	2018 Sep 20& \Hbeta                      & 6600&  1.6 \\
	          & \Hbeta{} continuum          & 6600&  1.6 \\	 
		\hline
    \end{tabular}
\end{table}

\subsection{SCORPIO-2 spectroscopy}

\label{sec:spectra}

In order to check the calibration of the line flux ratios measured by MaNGaL, we used two medium-resolution ($FWHM\sim$4.5~\AA) slit spectra presented by Boley et~al. (in prep.), obtained with the SCORPIO-2 instrument \citep{scorpio2} on the 6-m BTA telescope of SAO RAS in February, 2019.  The spectra cover two portions of the S235 nebula along the 6\arcmin{} slit, illustrated in Fig.~\ref{fig:MANGALres}, and the H$\alpha$, H$\beta$, \SII~6716, 6731~\AA{} lines are spectrally resolved, with an estimated absolute flux calibration accurate to about 5\%. We refer to the work of Boley et~al. for a full description of these observations and their reduction.

For each wavelength observed with MaNGaL, we matched the spatial resolution of the images (2.1\arcsec{}) to that of the SCORPIO-2 spectra (2.6\arcsec{}) by simple Gaussian convolution.  Next, we created spatial profiles of the emission in each line from the MaNGaL observations by integrating along the width of the slit (1\arcsec{}) at each position along the length of the SCORPIO-2 slit.  Finally, because of the lower SNR in the slit spectra of the fainter regions of the nebula, we applied 9-pixel (3.1\arcsec{}) boxcar smoothing to the spatial profiles (from both MaNGaL and SCORPIO-2).

\section{Methods}

\subsection{Properties of the ionized gas}\label{sec:ne}

Electron density \nelec{} was determined from the ratio of \SII{} lines  $\lambda 6716/ \lambda 6731$. Because the wavelength difference between the doublet lines is comparable to the MaNGaL FPI instrumental profile, a correction must be applied. For a Lorentzian profile, which is a good approximation of FPI instrumental contour \citep[][and references therein]{MoiseevEgorov2008}, the real ratio $r_{\rm real}$ of two lines separated by $\Delta\lambda$ is related to the observed ratio $r_{\rm obs}$ by the relation
\begin{equation}
    r_{\rm real} = \frac{1-C \cdot r_{\rm obs}}{r_{\rm obs} - C},
	\label{eq:rread}
\end{equation}
where $C = 1+ \left( \frac{2\Delta \lambda}{FWHM} \right)^2$ characterises the Lorentzian profile.

The observed MaNGaL map of \SII$\lambda 6716/ \lambda 6731$ was corrected according to Eq.~\eqref{eq:rread}.  Finally, we compared the spatial profiles of the \SII{} line ratio along the SCORPIO-2 slit with the same locations in the MaNGaL data using the procedure described in Sec.~\ref{sec:spectra}.  We found that the \SII$\lambda 6716/ \lambda 6731$ and \Halpha/\Hbeta{} values measured by MaNGaL must be further multiplied by a constant factor of 1.186 and 0.798 respectively to bring them in line with the flux-calibrated slit spectra.  This secondary correction may be due to several factors, including deviations of the real on-sky instrumental profile wings from the purely Lorentzian profile of the calibration lamp spectrum.

The value of \nelec{} was then calculated at each pixel with S/N level $>3$ from the corrected ratio of the \SII{} lines using Eqs.~(3) and (4) of \citet{Proxauf14}, for a mean electron temperature $T_{\rm e} = 7280$~K (Boley et~al., in prep). If the variations of electron temperature throughout the nebula are assumed to be on the order of $\sim170$~K \citep[][and references therein]{Esteban18}, the effect on the \nelec{} values derived from the relations of \citet{Proxauf14} are less than 2.5\%. Therefore, we do not attempt to account for variations in \Telec{} when considering trends in \nelec{} throughout the nebula. 

We determined the interstellar extinction \AV{} using the observed H$\alpha$/H$\beta$ intensity ratio and the intrinsic ratio for Case~B conditions and the reddening law of \citet{Cardelli89}. By interpolating the values from Table~4.2 of \citet{Osterbrock06} in log-log space for $T_e=7280$~K, we find an intrinsic intensity ratio of $j_{\mathrm{H}\alpha}/j_{\mathrm{H}\beta}=2.95$.  For the ratio of total to selective extinction, we adopted a value of $R_V=3.0$ (Boley et~al., in prep.).

To study the geometry of the \hii{} region, we estimate its depth along the line of sight ($S$) for the each pixel using equation:
\begin{equation}
S = \frac{4\pi I_{\rm H_{\alpha}}} {h \nu_{\rm H_{\alpha}}} \frac{1}{n_{\rm e} n_{\rm H^+} \alpha_2}~{\rm cm},
\label{eq:emissionmeasure}
\end{equation}
where $\alpha_2 = 3.94\times{}10^{-14}$~cm$^3$~s$^{-1}$ is the Case~B hydrogen recombination coefficient, interpolated for an electron temperature of $T_{\rm e}=7280$~K \citep[Table 4.2, ][]{Osterbrock06}, $I_{\rm H_{\alpha}}$ is from original calibrated MaNGaL files in ergs~cm$^{-2}$sec$^{-1}$arcsec$^{-2}$ transferred to ergs~cm$^{-2}$sec$^{-1}$sr$^{-1}$. We assume that the physical conditions ($n_{\rm e}$ and $T_{\rm e}$) do not vary over the line of sight, and ignore the ionization of all atoms other than hydrogen, i.e. $n_{\rm e} = n_{\rm (H^+)}$. While the physical conditions might not be truly uniform along the line of sight, we find below that the derived values of $S$ are reasonable and consistent with the overall structure of the \hii{} region seen in the distribution of \AV{} and \nelec{}. Regarding the ionization conditions, the abundance of ionized helium, which produces additional free electrons in the \hii{} region, is about 100 times less than that of H$^+$. The abundance of other ionized elements is even smaller.

\subsection{Dust temperature and column density}
\label{sec:dusttemp}

In order to estimate dust temperature and column density, we used far-infrared emission maps at 90, 140, and 160~$\mu$m \citep[WIDE-S, WIDE-L and N160 bands, respectively,][]{Kawada07}, collected by the Far-Infrared Surveyor (FIS) instrument on the {\it AKARI} satellite \citep[][]{Murakami07,Kaneda07} during the AKARI Far-infrared All-Sky Survey \citep{Doi15, Takita15}. According to \citet{Kawada07}, the average spatial resolution (FWHM) of the instrument at the IR bands listed above is 39, 58, and 61\arcsec, respectively. Later measurements produced somewhat greater FWHM values and revealed the elliptical shape of the PSF, which is more elongated at shorter wavelengths \citep{Takita15,2019PASJ...71....5U}. According to  \cite{2019PASJ...71....5U}, the in-scan FWHMs are $103.8\pm5.1''$, $104.2\pm7.9''$, and $85.5\pm8.3''$ for 90, 140, and 160 $\mu$m, respectively, while the cross-scan FWHMs for these same wavelengths are $68.5\pm3.1''$, $77.9\pm7.0''$, and $73.3\pm5.8''$. Based on these estimates, we conclude that the spatial resolutions of the maps at the wavelengths used are similar, so we have not attempted to convolve them to the same resolution. Dust temperatures and column densities were computed with the modified black body approach, adopting $\beta=-1.59$ and an opacity at 850~$\mu$m of
0.5 cm$^2$ g$^{-1}$ \citep{2014A&A...571A..11P}.

We downloaded the AKARI data from the IRSA archive\footnote{\url{https://irsa.ipac.caltech.edu}} and regridded them to the astrometric grid of the 90~$\mu$m image with a pixel size of 15\arcsec. We used the standard HASTROM IDL procedure\footnote{\url{https://idlastro.gsfc.nasa.gov}} with the default bilinear interpolation. Dust temperatures and column densities were computed with the modified black body approach, adopting various prescriptions for dust opacities (see below). Having obtained the dust temperature \Tdust{} from the SED fitting, we applied the general approach described by \citet{1983QJRAS..24..267H} and used a standard dust-to-gas mass ratio 1/100 to obtain the hydrogen column density map.
 
To estimate the radiation field in units of the Habing field \citep[$G_0$][]{1968BAN....19..421H}, we use equation (5.44) from \citet{tielensbook} for a grain radius of 0.1~$\mu$m:
\begin{equation}\label{eq:G0}
    T_{\rm sil} \simeq 50 \left( \frac{1\mu {\rm m}}{a} \right)^{0.06} \left( \frac{G_{\rm 0}}{10^4} \right)^{1/6}.
\end{equation}
We use 0.1~$\mu$m as a reference size of a typical interstellar grain. This choice is motivated the fact that the peak of the grain size distribution is close to 0.1--0.3~$\mu$m \citep[e. g.][]{2001ApJ...548..296W}. Additionally, the interstellar grain size distribution, recalculated in terms of mass fraction, has a peak near 0.1~$\mu$m \citep{1994ApJ...422..164K}.

\section{Results of the observations with MaNGaL}\label{sec:res}

Images of the emission in the \Halpha{}, \Hbeta{}, \SII, and \NII{} lines are shown in Fig.~\ref{fig:MANGALres}. Only pixels with S/N higher than 3 are shown. In Appendix~\ref{sec_errors} (supplementary material available online) we show the maps of signal to noise ratios for each. The images of \Halpha{} and \Hbeta{} show a bright central part (`main' part) around the ionizing star, with a sharp edge to the north and west, and diffuse emission to the south. The peaks of the \Halpha{} and \Hbeta{} emission are about 60\arcsec{} to the north-east of the ionizing star (0.5~pc in the plane of the sky). The area of the brightest \Halpha{} and \Hbeta{} emission is marginally coincident with the 1.4~GHz radio emission peak. The radio continuum emission is also present in the diffuse, southern part of the optical nebula. 

In contrast to the bright \Halpha{} emission, \SII{} lines were only detected in the main part of the nebula with S/N $> 3$. The peak of the \SII{} emission is coincident with the peak of \Halpha{} and \Hbeta{} emission. Several arc-like filaments are visible in the \SII{} maps to the north-east from the ionizing star. There is also a bright, separated filamentary structure in the north-west part of the main nebula detected in all five lines.

\begin{figure*}
	\includegraphics[width=\columnwidth]{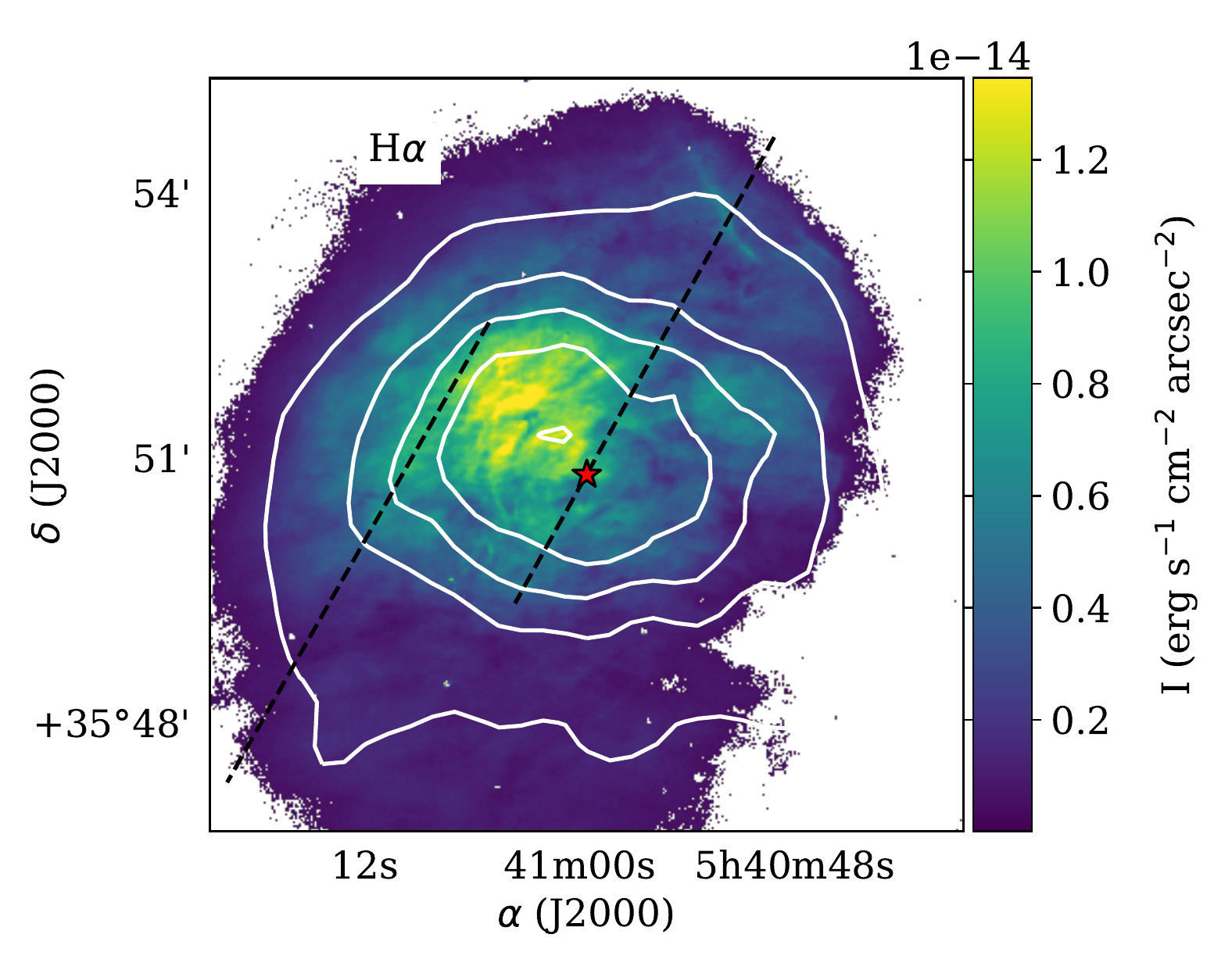}
	\includegraphics[width=\columnwidth]{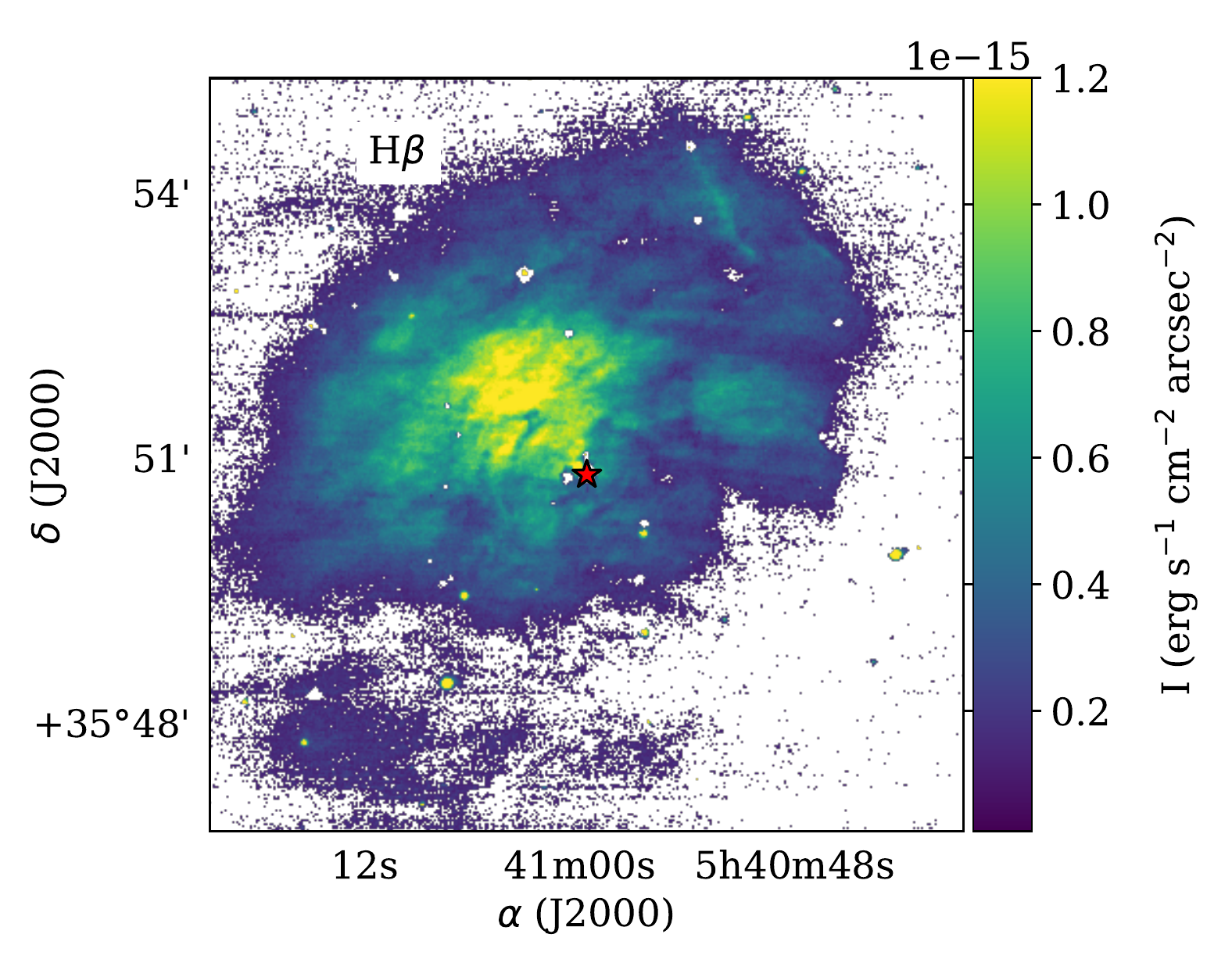}
	\includegraphics[width=\columnwidth]{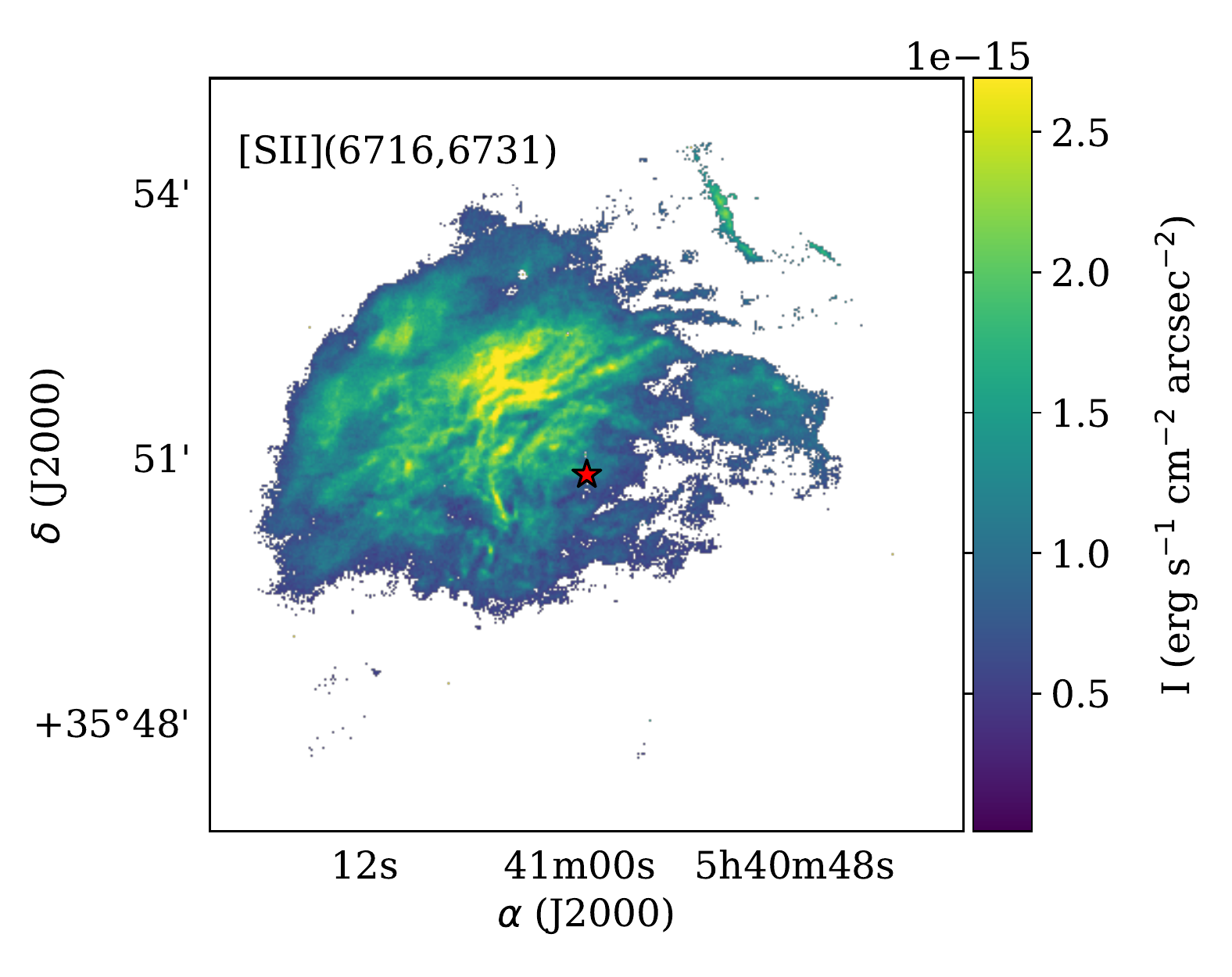}
	\includegraphics[width=\columnwidth]{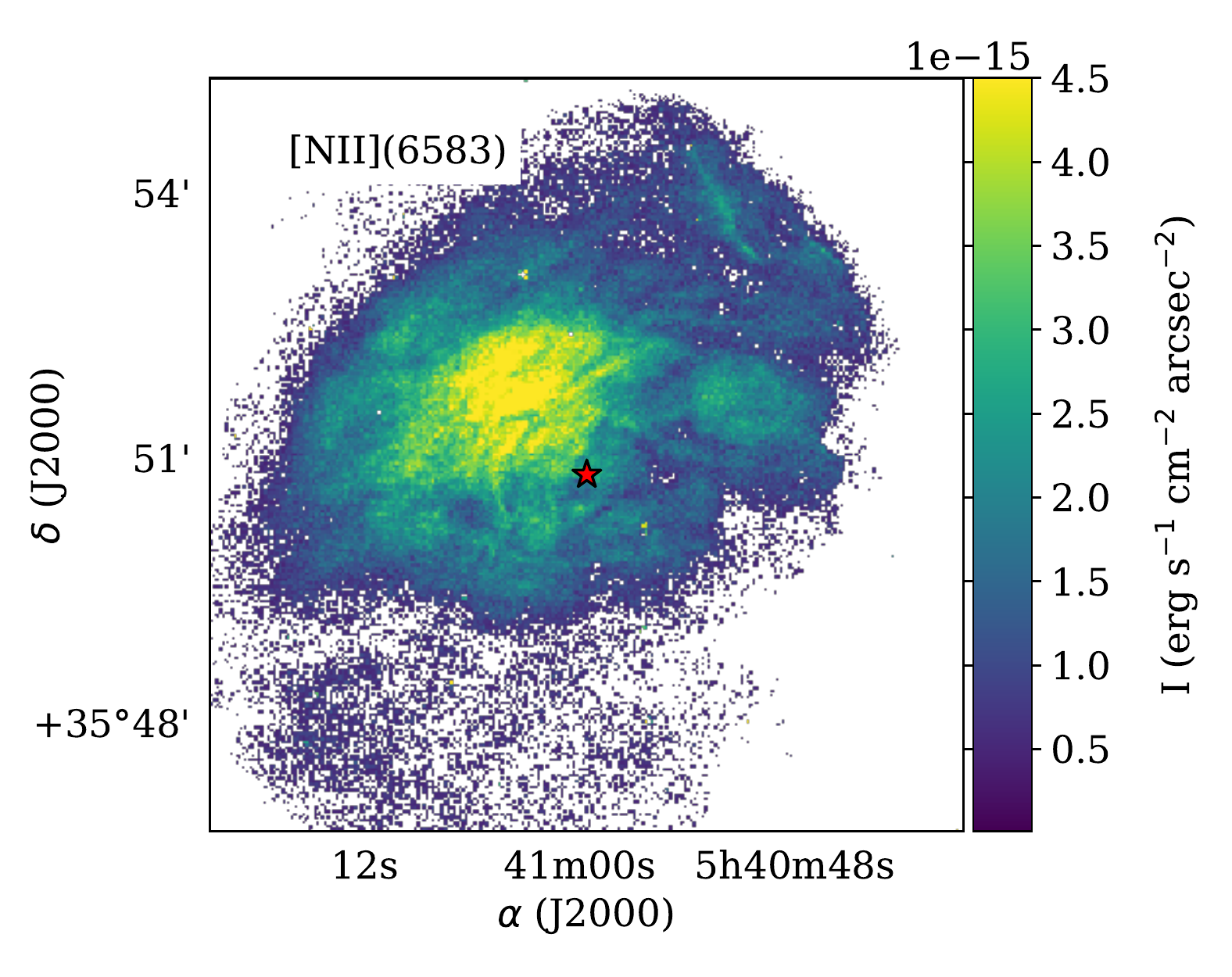}\\
	
    \caption{MaNGaL images of S235. Only pixels with S/N ratio $>3$ are shown here. The NVSS 1.4~GHz image of S235 \citep{Condon98} is superimposed on the \Halpha{} image with white contours, which are evenly spaced from 2.25~mJy/beam ($5\sigma$ level) to 123~mJy/beam, and the dashed lines show the slit positions of the SCORPIO-2 spectra. The location of the ionizing star is shown in each frame as a red star.}
    \label{fig:MANGALres}
\end{figure*}

\begin{figure*}
	\includegraphics[width=\columnwidth]{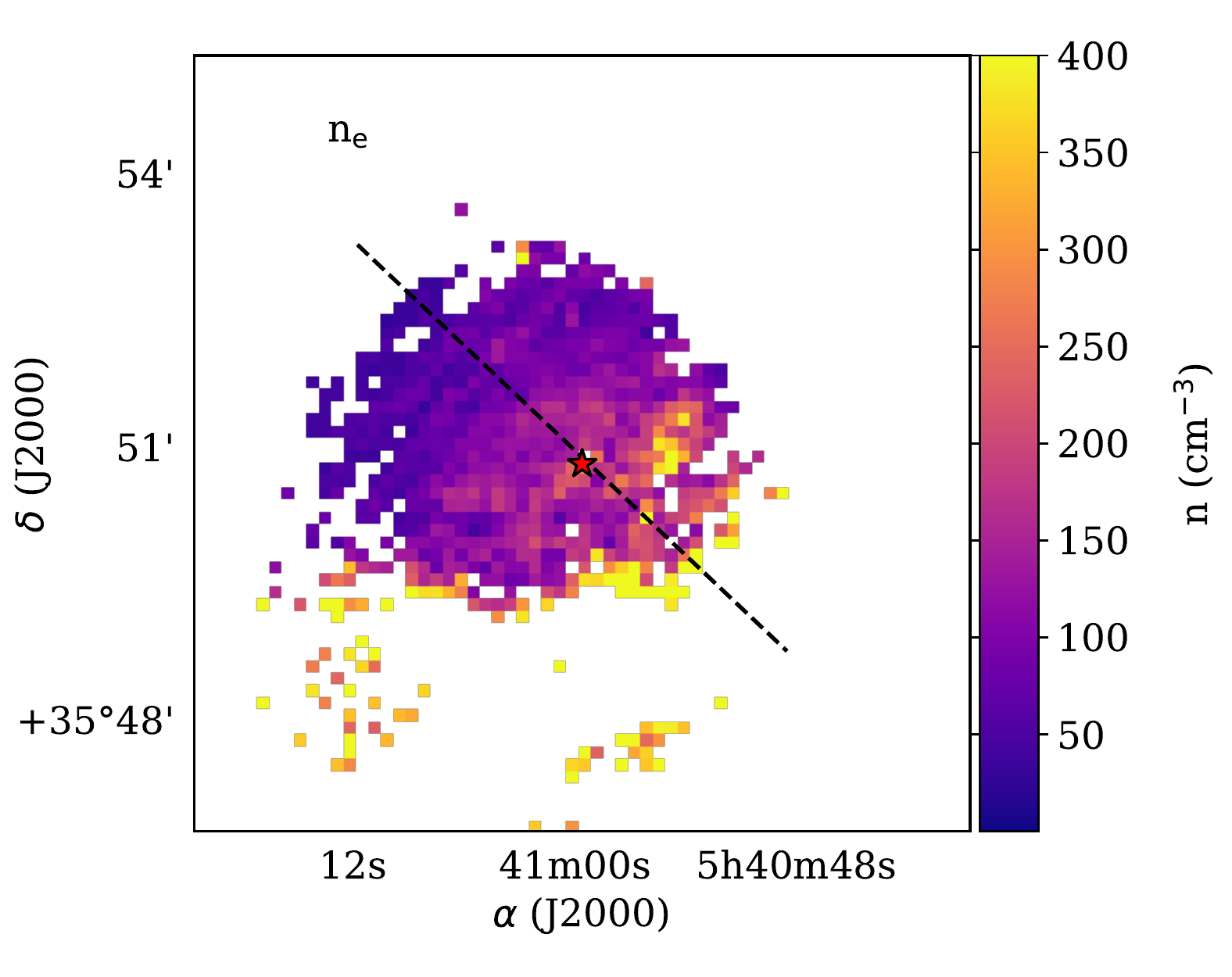}
	\includegraphics[width=\columnwidth]{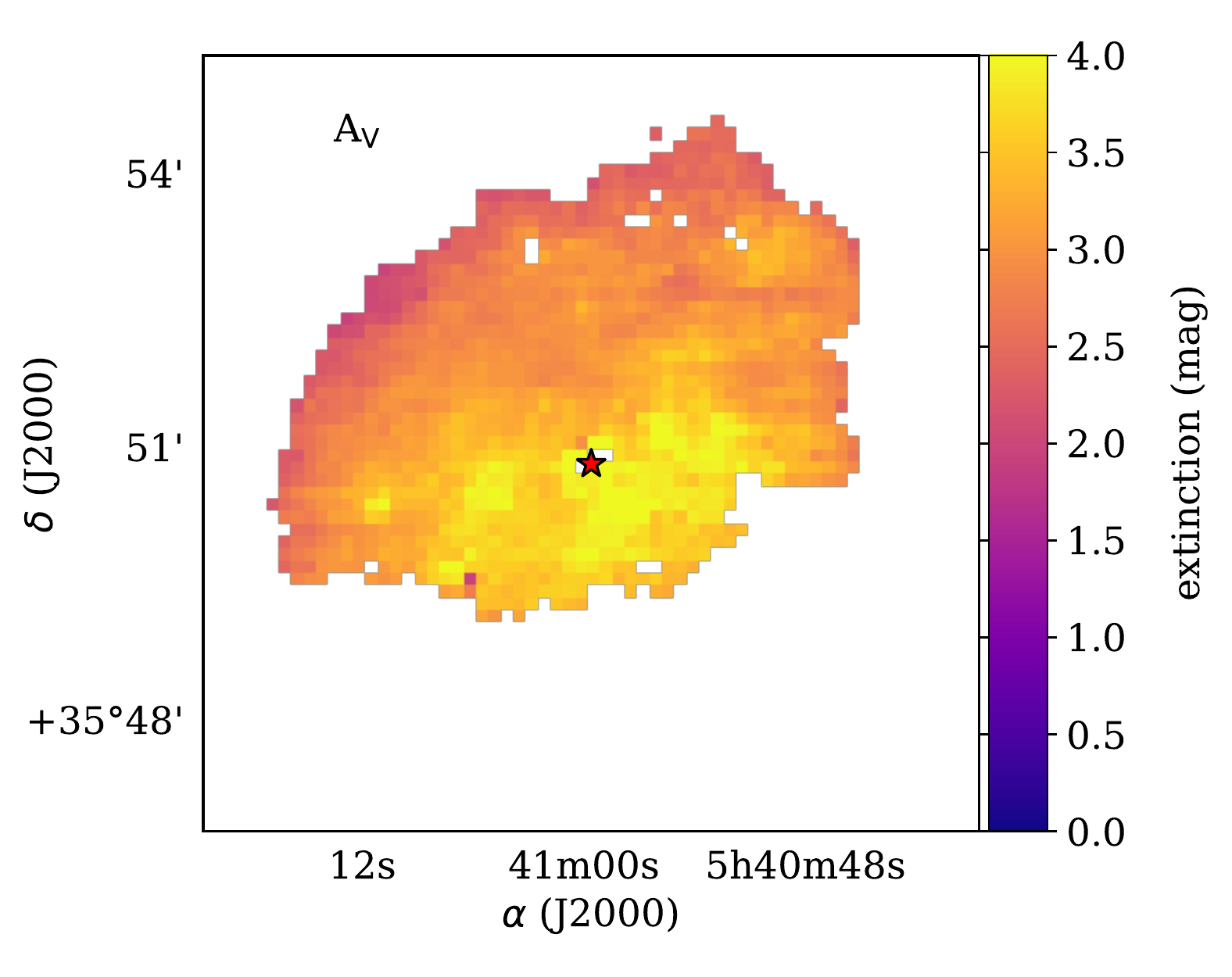}\\
	\includegraphics[width=\columnwidth]{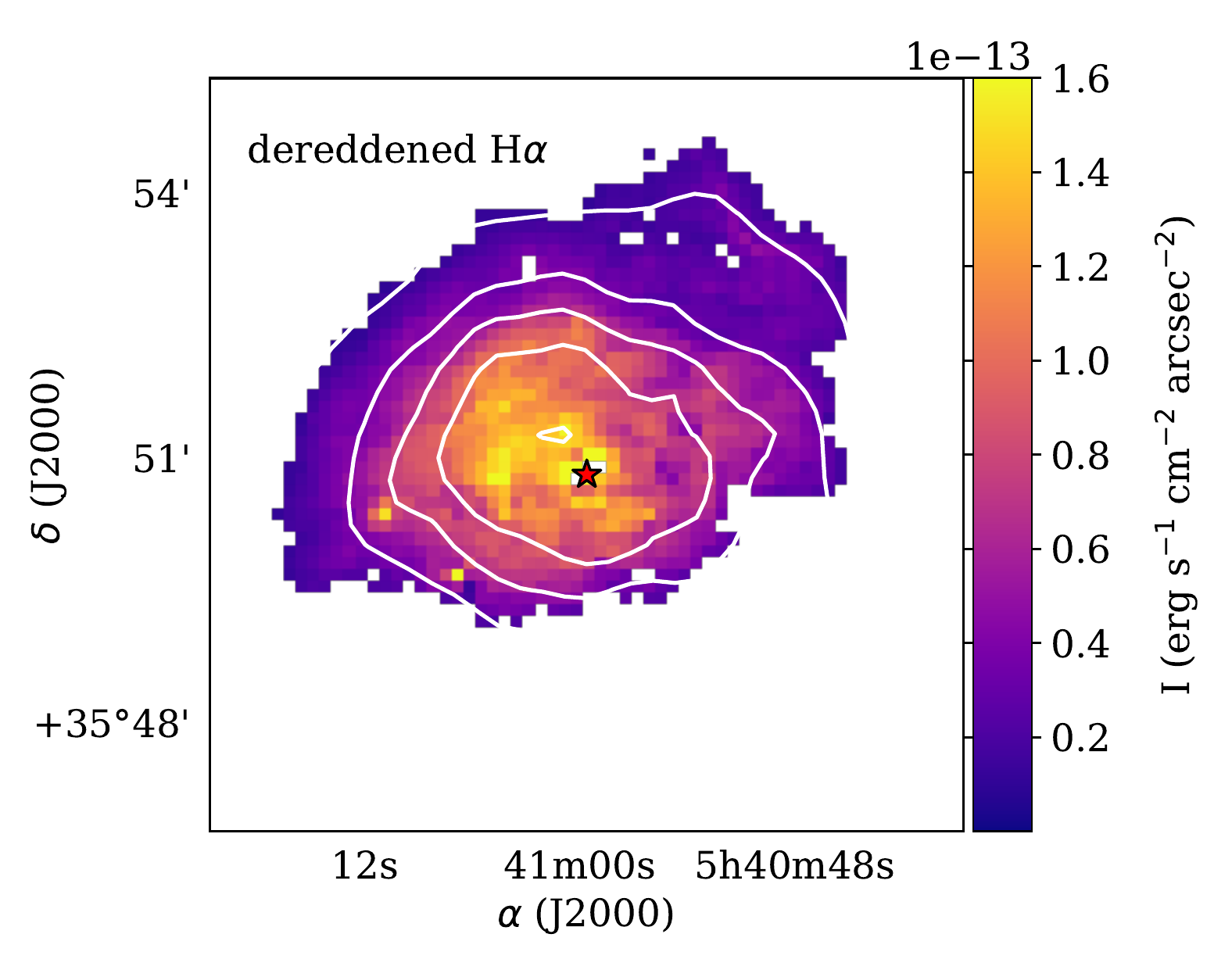}
	\includegraphics[width=\columnwidth]{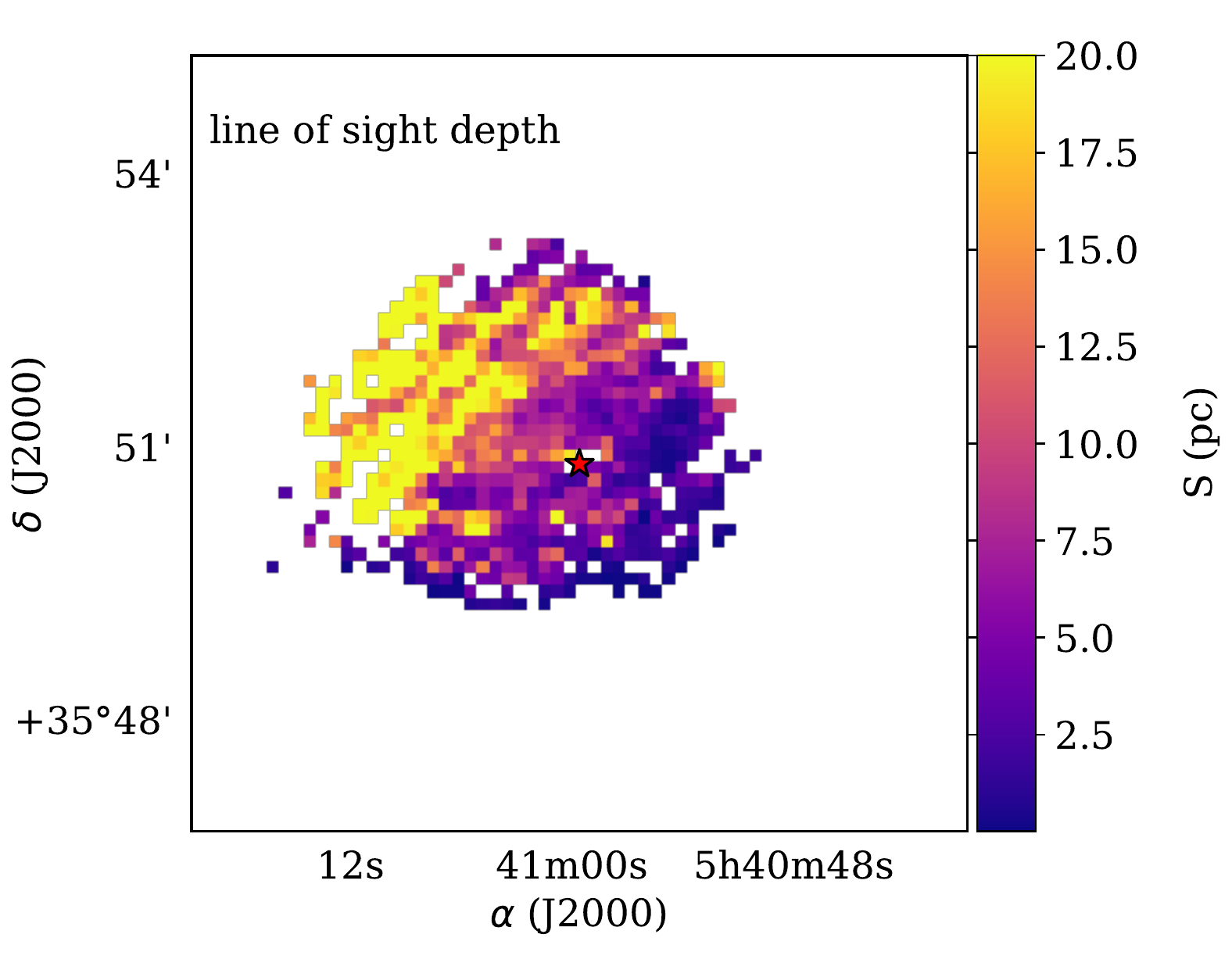}
	\caption{Rebinned electron density (top left) and \AV{} (top right),  dereddened \Halpha{} image (bottom left) and line-of-sight depth (bottom right) images of S235. Only pixels with S/N ratio $> 3$ are shown. The dashed line in the top left panel shows the location of the cut of the \nelec{} value presented in Fig.~\ref{fig:eldnsitycut}. }
    \label{fig:eldens}
\end{figure*}

\begin{figure}
\includegraphics[width=1.05\columnwidth]{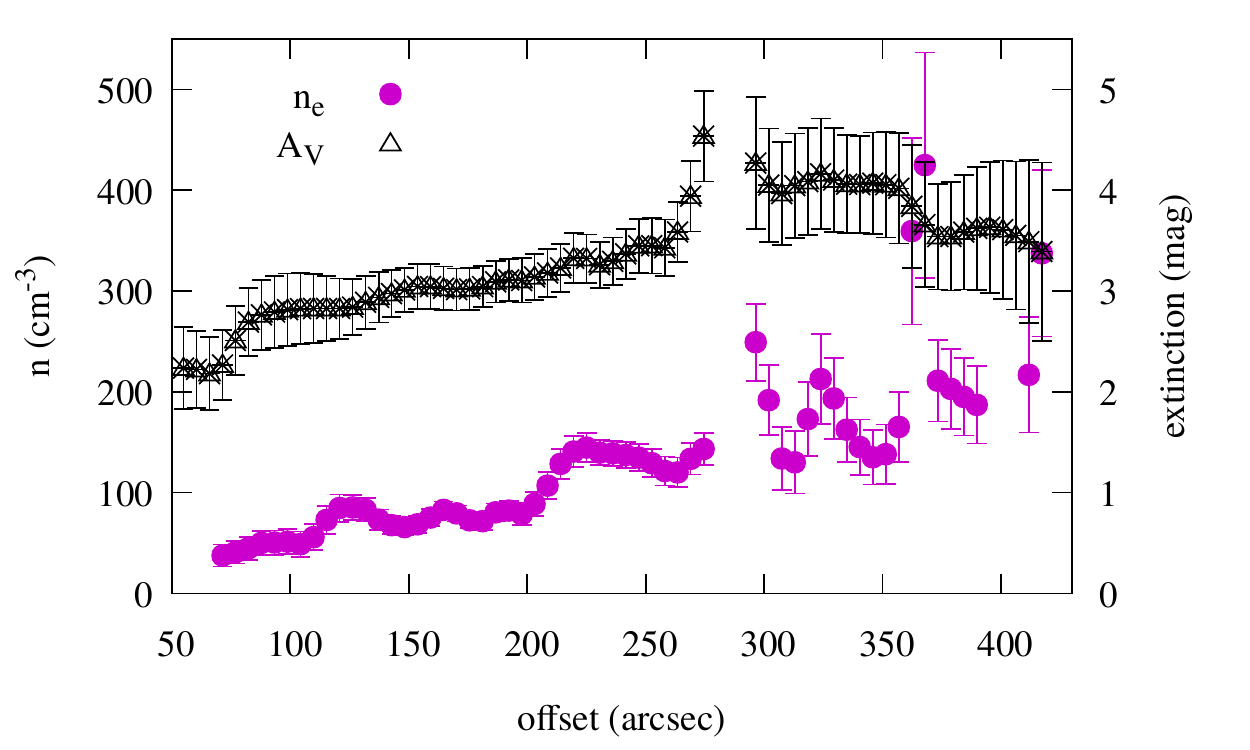}
\caption{ Cuts of \nelec{} (violet circles) and \AV{} (black triangles) along the north-east/south-west direction. The position of the cut on the plane of the sky is shown in Fig.~\ref{fig:eldens}. The empty region on the cut corresponds to the position of the ionizing star and surrounding pixels.}
\label{fig:eldnsitycut}
\end{figure}

In Fig.~\ref{fig:eldens}, we show the spatial distribution of \nelec{} throughout the nebula. As we are interested in large-scale trends, and in order to increase the S/N ratio, the images were rebinned to a pixel size 16 times larger (8.1\arcsec) than the original pixel shown in Fig.~\ref{fig:MANGALres}. We show the resulting S/N ratio of the \nelec{} distribution after this procedure in Appendix~\ref{sec_errors}. We find a density gradient from the north-east to the south-west part of the nebula, from $\sim20-30$~cm$^{-3}$ to more than 300~cm$^{-3}$, with a median value of 96~cm$^{-3}$. We show a north-east/south-west spatial cut of the \nelec{} value and its uncertainty in Fig.~\ref{fig:eldnsitycut}. Both \nelec{} value and its uncertainty grow to the south-west. The gap in the cut at the position of the ionizing star is due to artefacts remaining from subtracting the bright stellar continuum at this location. The ionizing star is projected onto a region with $n_{\rm e} > 150$~cm$^{-3}$.   Due to the faintness of the \SII{} lines in the south and south-west regions, the value of \nelec{} is determined with greater uncertainty, and at some locations cannot be determined at all. The bright north-west filament visible in Fig.~\ref{fig:MANGALres} is not distinguishable on the \nelec{} map.

We performed a visual comparison of the \nelec{} map from Fig.~\ref{fig:eldens} with the spatial distribution of hydrogen and carbon radio recombination lines (RRLs and CRRLs, respectively) given by \cite{Anderson19}, shown in their Fig.~7. Due to the square-root dependence on density, the RRLs are the best tracers of \nelec{} in the \hii{} region, but CRRLs trace \nelec{} in the surrounding PDR. We see that the peak of the RRL emission corresponds to the direction of the ionizing star, but the peak of the CRRL emission is shifted to the south-west of the star and coincides with the region of the maximum \AV{} found with MaNGaL. Hence, the densest part of the \hii{} region is adjacent to the densest part of the PDR around S235.

 The spatial distribution of the optical extinction, expressed in terms of \AV{}, is also non-uniform in S235 (see Fig.~\ref{fig:eldens}). There is a gradient from $A_{\rm V}=1.5-2$~mag in the north-east part of the nebula to $A_{\rm V}>4$~mag in the south-west part. The \AV{} value in the direction of the ionizing star is $\approx 4$~mag, consistent with the value found by \citet{Straizys10}. The typical uncertainty of the \AV{} value, which we calculated using a bootstrapping procedure by varying the \Halpha{} and \Hbeta{} intensities within their respective uncertainties, is about 0.2~mag, which is less than 6\% of the minimum \AV{} value found in the north-west part of the nebula. The map of the signal to noise ratio for \AV{} is shown in Appendix~\ref{sec_errors}. Comparing the \nelec{} and \AV{} maps, we find that the direction of the gradients is approximately the same, with higher \nelec{} corresponding to higher \AV{}. The ionizing star is projected on the transitional region between the dense part of the \hii{} region more deeply embedded in the surrounding neutral material and the more rarefied part. The transition from the less-dense and less-obscured part of the \hii{} region to the denser part is visible in Fig.~\ref{fig:eldnsitycut}. The physical properties, as well as their uncertainties, change when crossing the star position.

The dereddened and rebinned \Halpha{} image is shown in Fig.~\ref{fig:eldens}. After accounting for the foreground reddening, it can be seen that maximum of the \Halpha{} emission coincides with the maximum of the 1.4~GHz NVSS image. The uncertainty of the dereddened \Halpha{} flux reaches 50\% in the south-west part of the nebula, where the \AV{} value is highest, but not more than 10--20\% in other parts of the nebula.

We evaluated the extent of the \hii{} region along the line of sight $S$ using Eq.~\eqref{eq:emissionmeasure} and the distribution of the \nelec{} value, and  found that it varies within the range of $2 < S < 20$~pc in S235. The linear size of the nebula in the plane of the sky, corresponding to 5\arcmin{} at 1.65~kpc, is 2.4~pc. Thus, according to the optical data, the nebula is more extended along the line of sight. If we use the emission measure from the radio recombination lines of $EM=3.7\cdot10^4$~pc~cm$^{-6}$~\citep{1978ApJ...224..437S} with the median value of \nelec, we obtain 4.0~pc; i.e. our estimate of the extent from the optical recombination lines is in agreement with the radio-based value within a factor of 2--5, depending on the position in the nebula.

The value of $S$ shows a gradient opposite in sign compared with both the \nelec{} and \AV{} gradients, i.e. it increases from the south-west to north-east. The depth is about 2--7~pc at the location of the central star. Therefore, there is a more or less spherical/ellipsoidal volume of ionized gas to the south-west of the star. In the north-east part of the nebula, we find gas that is less dense, less obscured, and more extended along the line of sight. This region represents a location where the gas can more easily expand from the \hii{} region into the direction of the observer. Indeed, \citet{Anderson19} also showed a difference between the radial velocities of the hydrogen and carbon RRLs (--23 and --20 \kms, respectively) accompanying the expansion.

\section{Analysis of dust emission}

\begin{figure}
	\includegraphics[width=\columnwidth]{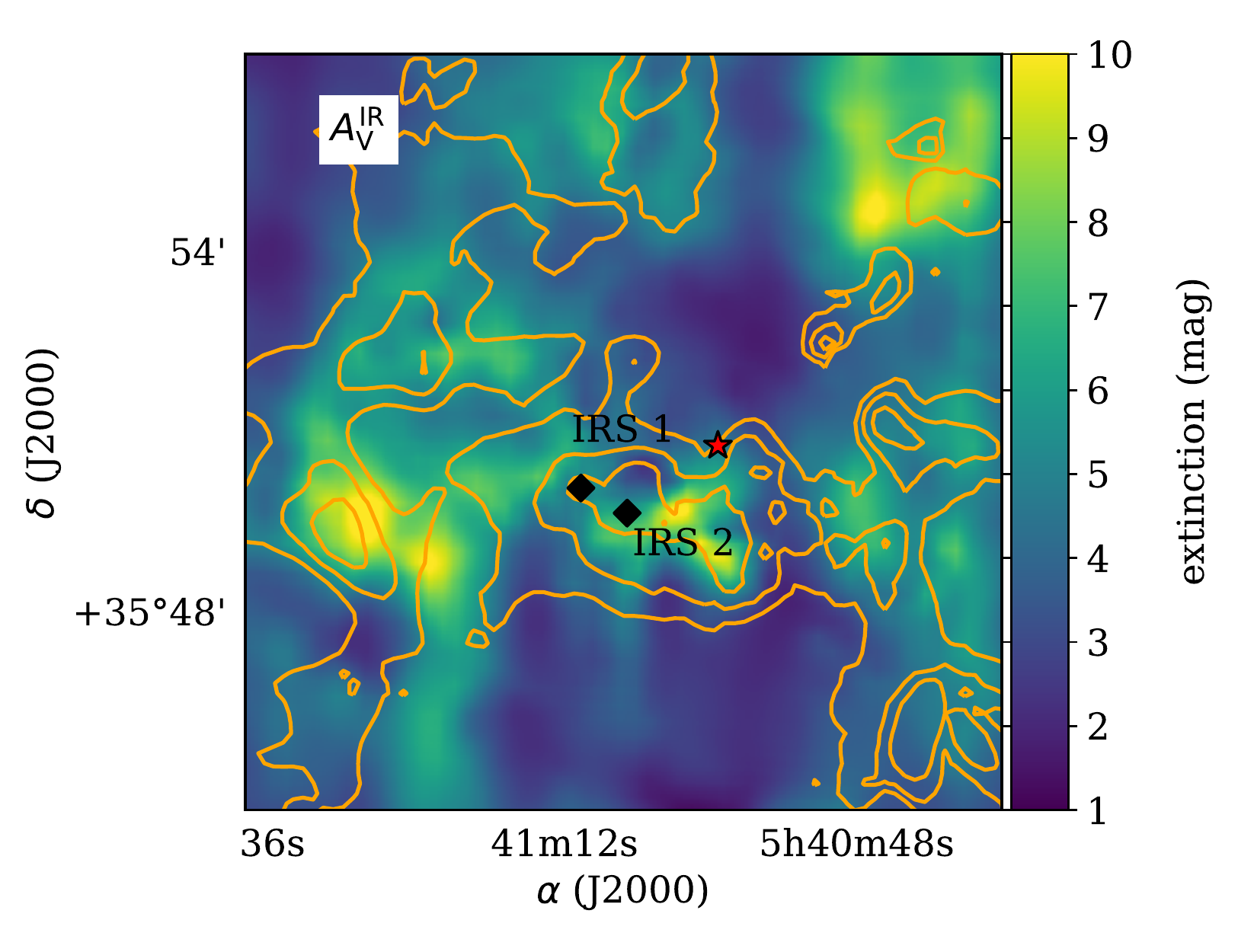}\\
	\includegraphics[width=\columnwidth]{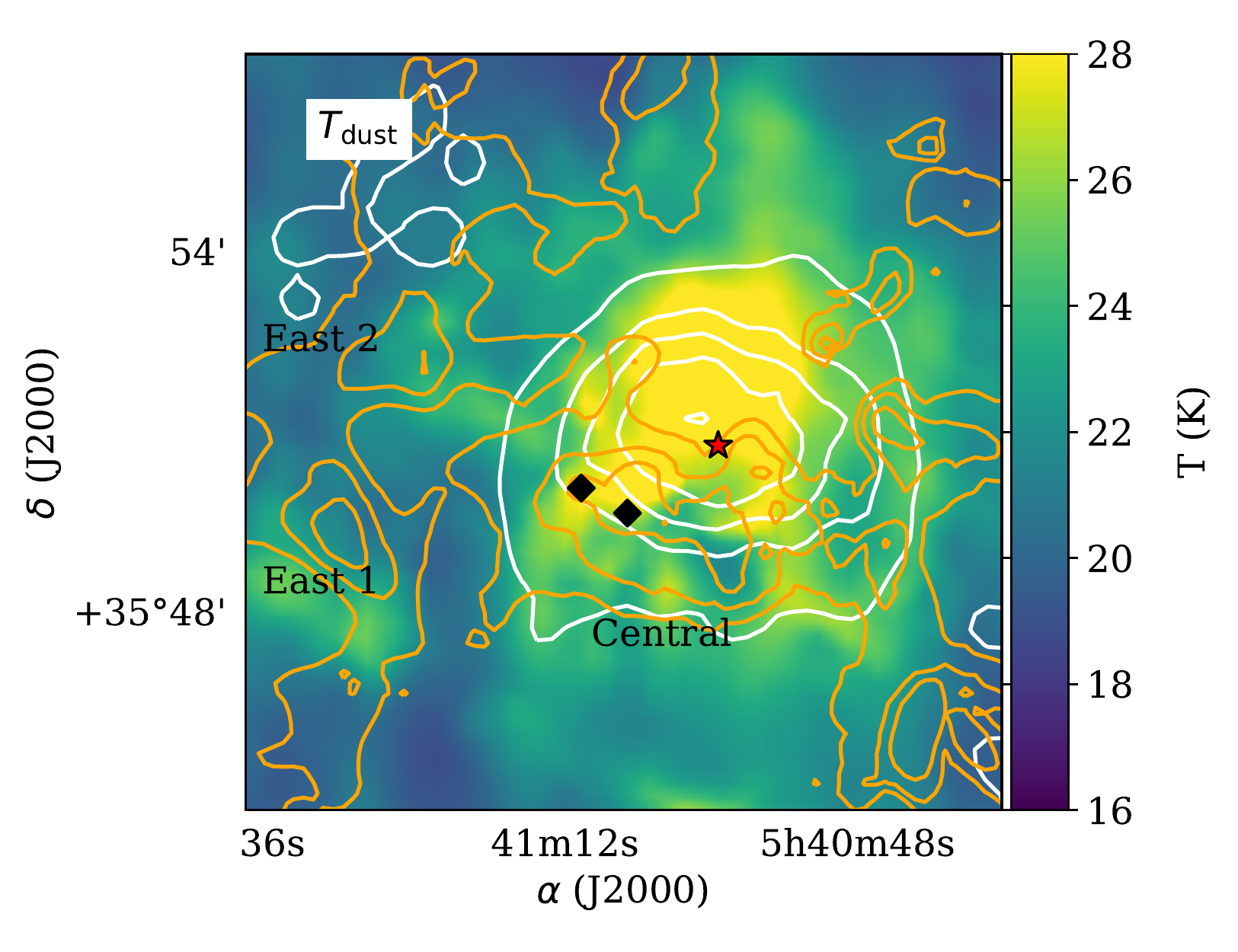}\\
	\includegraphics[width=\columnwidth]{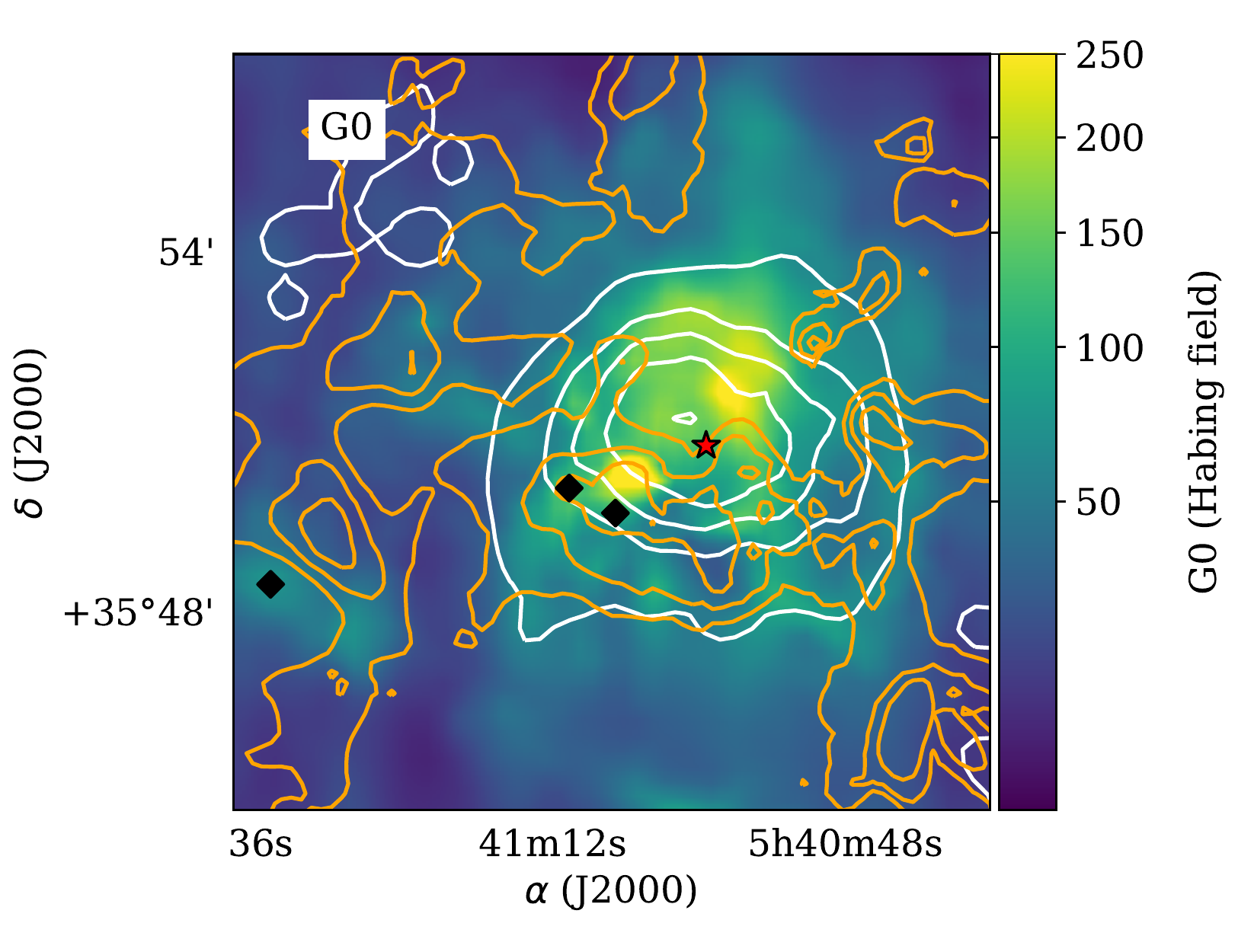}\\
	\caption{Top: dust extinction, middle: dust temperature, bottom: UV field in Habing units. NVSS radio continuum emission \citep{Condon98} is shown with white contours. The ionizing star is shown as the red star symbol, and the infrared sources IRS~1 and IRS~2 \citep{Evans81a} are shown as black diamonds. The map of $N({\rm H_2})$ from \citet{Bieging16} is shown with orange contours, and dense clumps with embedded young stellar clusters from \citet{Kirsanova08} are designated. The contours, in terms of visual extinction, correspond to 5, 10 and 15~mag. The black diamond shown on the bottom panel indicates the position a possible illuminating source for S235~East~1~E PDR (see text).}
    \label{fig:tdust}
\end{figure}

There are three different types of column density indicators which can be applied to this region. One is \AV{} derived from the Hydrogen Balmer lines in our MaNGaL observations. Another, $A^{\rm IR}_{\rm V}$, may be obtained from spectral fitting of the dust continuum emission, as described in Sec.~\ref{sec:dusttemp} (see also Fig.~\ref{fig:tdust}, top). Finally, $A^{\rm CO}_{V}$ may be derived from the CO observations of \citet{Bieging16}, assuming a standard LTE approximation \citep{2015PASP..127..266M} and relative abundance for CO of $1.2\cdot10^{-4}$ (see, for example, the work by \citet{Wakelam2008}; this value is also in agreement with the latest calculation of carbon abundance in the region made with the CO, C$^+$ and 13C$^+$ lines by Kirsanova et~al., in prep.). In a simple spherical-shell approximation, and for a line of sight passing through the shell centre, the first value (\AV{}) contains contributions from the ionized region and the front wall. The second value ($A^{\rm IR}_{\rm V}$) also contains contributions from the ionized region, as well as from both the front and rear walls. Finally, the third value ($A^{\rm CO}_{V}$) is presumably sensitive to the (molecular) material contained in the front and rear walls.

In this picture, $A^{\rm IR}_{V}$ should be the largest (provided the dust temperature does not vary along the line of sight), which is indeed what we see in general. In the region observed with MaNGaL, the median extinction values are 2.7, 4.0, and 2.9~mag for \AV, $A^{\rm IR}_{\rm V}$, and $A^{\rm CO}_{\rm V}$, respectively. However, this relation breaks down significantly at certain locations. Specifically, $A^{\rm IR}_{\rm V}< A_{\rm V}$ at nearly all locations to the north of the ionizing star. Also, as the observer looks through the dense molecular clump southward of the ionizing star, $A^{\rm IR}_{\rm V}< A^{\rm CO}_{\rm V}$. This behaviour is somewhat unexpected, and may be caused by the choice of the opacity prescription. The Planck-based opacity parameters that we have utilised are more appropriate for the diffuse medium, while the gas in the studied region has a much higher density. We also tried other opacity prescriptions \citep[e.g.][]{1994A&A...291..943O}, and they do not improve the situation. The CO-based extinction is also somewhat uncertain, as its value scales linearly with the adopted CO abundance, and this value is also prone to some variations. Overall, we conclude that extinctions based on {\em AKARI} data, as well as the extinctions based on the CO data, are uncertain up to a factor of two, and should therefore be used as a relative measure of column density, rather than absolute. We note, however, that the estimated dust temperature does not depend strongly on the adopted opacity.

The $A^{\rm IR}_{\rm V}$ varies from 2~mag to the north of the ionizing star BD+35$^\circ$1201 within the \hii{} region, to more than 10~mag in the densest regions, which coincide with the peaks of the CO($2-1$) emission \citep{Bieging16}. If we consider only the area outlined by the NVSS contours, we find that the column density traced by both $A^{\rm IR}_{\rm V}$ and $A^{\rm CO}_{\rm V}$ increase from north to south, so that the directions of their gradients are not the same as the direction of the \AV{} gradient estimated with the hydrogen recombination lines. This implies that the front and rear walls of neutral material of the \hii{} region are inhomogeneous, and contain different amounts of absorbing material.

Comparing the extinction values from Figs.~\ref{fig:eldens} (top right) and \ref{fig:tdust} (top), we find that while $A^{\rm IR}_{\rm V}$ is about 1.5 times as high as \AV{} in the direction of BD+35$^\circ$1201, the \AV{} value does not vary much as we look through the dense region with an embedded young stellar cluster (S235~Central; \citealt{Kirsanova08}) to the south of the ionizing star, where $A^{\rm IR}_{\rm V}$ is significantly higher than \AV. This implies that the cluster and the dense clump are situated in the rear wall of the \hii{} region. This conclusion is in agreement with findings by \citet{Anderson19}, who compared the radial velocities of carbon and hydrogen RRLs with the velocities of molecular gas in the same region. They found that the ionized gas flows in the direction of the observer, and that the neutral material forms a semi-envelope around the \hii{} region from the rear and two side walls. In the present study, we show that the \hii{} region has an inhomogeneous front wall, with a smaller column of neutral material than the rear wall.

The dust temperature derived from the SEDs (Fig.~\ref{fig:tdust}, middle) represents the average value of \Tdust{} along the line of sight and varies from 18 to 30~K across the face of the \hii{} region and the surrounding PDR, with a minor peak to the south-east of the ionizing star. Comparison of the CO($2-1$) emission and corresponding $N({\rm H_2})$ with the spatial distribution of the \Tdust{} values shows that the peaks of the CO($2-1$) in direction of the stellar clusters S235~East~1 and East~2 \citep{Kirsanova08} to the east of the \hii{} region coincide with the regions with \Tdust{} < 20~K. We do not find decreasing \Tdust{} in the direction of S235~Central, probably due to the projection effect of the warm foreground dust heated by the BD+35$^\circ$1201 star.

The typical value of FUV radiation field is $G_{\rm 0} \approx 200$ in Habing units in the direction of the \hii{} region (Fig.~\ref{fig:tdust}, bottom). The maximum value of $G_{\rm 0}$ does not correspond to the position of the ionizing star or to the bright infrared sources IRS~1 or IRS~2 found by \citet{Evans81a}, but rather with the local lows of the dust column density map to the north of the ionizing star, where the FUV emission is not absorbed by the dense neutral medium. The minimum FUV field in the direction of the S235~East~1 and East~2 clusters can be explained by the high gas density in the molecular clumps. The projected distance from the ionizing star to the western border of the dense gas ridges in the direction of the East~1 and East~2 clusters is $\approx6$\arcmin, which corresponds to 2.8~pc. The value of $G_{\rm 0}$ is $\approx 30-50$ in the direction of the western border of East~1 from the side of the ionizing star (S235~East~1~W below). 

We find $G_{\rm 0} \approx 100$, or 2-3 times higher, on the eastern border of East~1 (which we call S235~East~1~E below). The dense clump with the embedded cluster East~1 is probably irradiated from the east by another stellar source, visible, for example, on {\it WISE} or {\it Spitzer} images of the region at $\alpha$(J2000.0)=05$^h$41$^m$35.23$^s$, $\delta$(J2000.0)=35$^\circ$48\arcmin27.5\arcsec{} (the source is shown in the bottom panel of Fig.~\ref{fig:tdust}). The value of $G_{\rm 0}$ here is comparable with the Horsehead PDR, where $G_{\rm 0} \approx 100$~\citep{Zhou93}. The gas number density in the S235~East~1 dense clump reaches up to $2\times 10^4$~cm$^{-3}$ based on measurements of NH$_3$ lines by \citet{Kirsanova14}. Thus, the dense clump S235~East~1 is a two-sided dense PDR with different $G_{\rm 0}$ values at each side, making this region an interesting target for studies of PDR chemistry.

\section{Conclusions}

We present the first observations of a galactic \hii{} region with the optical tunable-filter photometer MaNGaL at the Zeiss-1000 telescope of SAO~RAS. The observations were done in \Halpha, \Hbeta, two [SII] lines at $\lambda 6716, 6731$~\AA{} and the \NII{} line at 6583~\AA{}. The distribution of absorbing material (in terms of \AV{}) was obtained using the \Halpha{} and \Hbeta{} images. The \SII{} lines were used to obtain the value of \nelec.

We conclude that optical emission of the \hii{} region is attenuated by neutral material with $A_{\rm V} \approx 2-4$~mag and a peak to the south-east from the ionizing star. The direction to the highest \AV{} coincides with the maximum detected electron density (up to $n_{\rm e} > 300$~cm$^{-3}$), with a median value 96~cm$^{-3}$.

The combination of the results of the optical observations with archive FIR data from the {\it AKARI} satellite allowed us to describe the 3D structure of the \hii{} region: we obtain a contribution of the front and rear walls to the total column density of neutral material. We find that the rear wall of the \hii{} region contains higher column of material than the front wall. This result agrees with recent study by \citet{Anderson19}, who found that the ionized gas of the \hii{} region expands in the direction of the observer. The thick rear wall does not allow to the ionized gas to expand to the direction from the observer. The extent of the \hii{} region along the line of sight varies from 2~pc in the south-west to more than 10~pc in the north-east direction.

We also estimated $T_{\rm dust}$ and the mean FUV field in terms of $G_0$ in S235 and the surrounding PDR, and found an interesting two-sided PDR in the dense clump S235~East~1, where $G_0=30-50$ and 100 in the western and eastern parts, respectively. This region is attractive for studies of dense PDR chemistry.

\section*{Acknowledgements}
We thank the anonymous referee for critique and suggestions which led to the improvement of this manuscript. We are also thankful to O.~V.~Egorov for fruitful discussions, as well as D.~V.~Oparin and A.~E.~Perepelitsyn for their assistance in observations.

Data processing and analysis of the observational material by MSK and PAB was supported by the Russian Science Foundation, grant number 18-72-10132.
The development of MaNGaL and the tunable-filter data reduction software were supported by the Russian Science Foundation, grant number 17-12-01335. This study is  based on observations conducted with the 1-m and 6-m telescopes of the Special Astrophysical Observatory of the Russian Academy of Sciences, carried out with the financial support of the Ministry of Science and Higher Education of the Russian Federation (including agreement No. 05.619.21.0016, project ID RFMEFI61919X0016), and also on observations with AKARI, a JAXA project with the participation of ESA. This work made use of data from the European Space Agency (ESA) mission
{\it Gaia} (\url{https://www.cosmos.esa.int/gaia}), processed by the {\it Gaia}
Data Processing and Analysis Consortium (DPAC,
\url{https://www.cosmos.esa.int/web/gaia/dpac/consortium}). Funding for the DPAC
was provided by national institutions, in particular the institutions 
participating in the {\it Gaia} Multilateral Agreement. This research made use of NASA's Astrophysics Data System Bibliographic Services, SIMBAD database,
operated at CDS, Strasbourg, France \citep{2000A&AS..143....9W},  Astropy,\footnote{http://www.astropy.org} a community-developed core Python package for Astronomy \citep{astropy:2013, astropy:2018}, Matplotlib \citep{Hunter:2007} and APLpy, an open-source plotting package for Python \citet{2012ascl.soft08017R}.

\section*{Data Availability}

The data underlying this article are available in Zenodo, at \url{https://doi.org/10.5281/zenodo.3902074}.

\bibliographystyle{mnras}
\bibliography{refs} 

\appendix

\section{Uncertainties of the obtained results}
\label{sec_errors}

In this section we provide the reader with maps of signal to noise ratio. We show the ratio for the obtained emission line maps with original binning. We obtain comparable signal to noise ratio for the \nelec{} and \AV{} values using 16 times larger pixel size than in the line emission maps.

\begin{figure*}
	\includegraphics[width=\columnwidth]{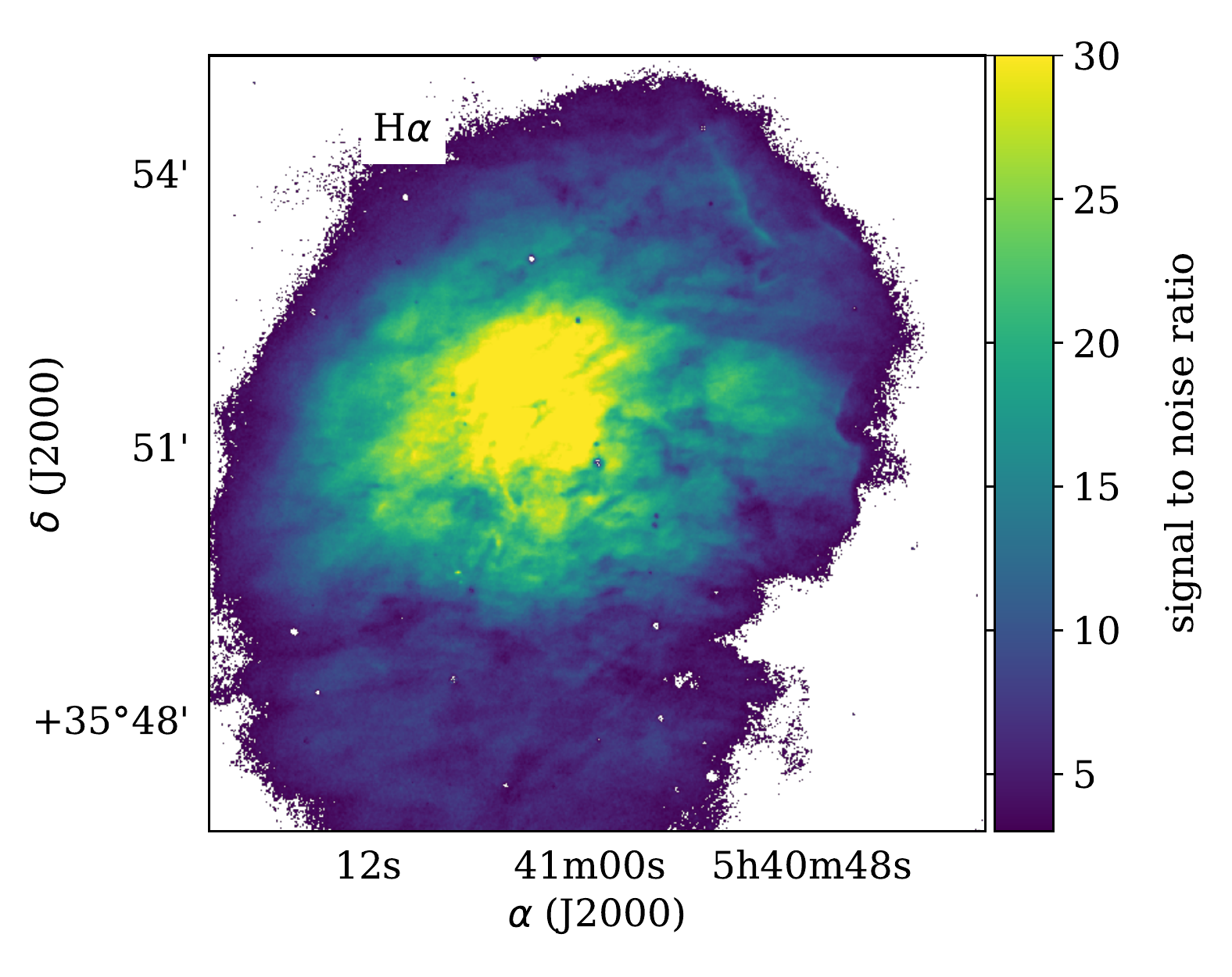}
	\includegraphics[width=\columnwidth]{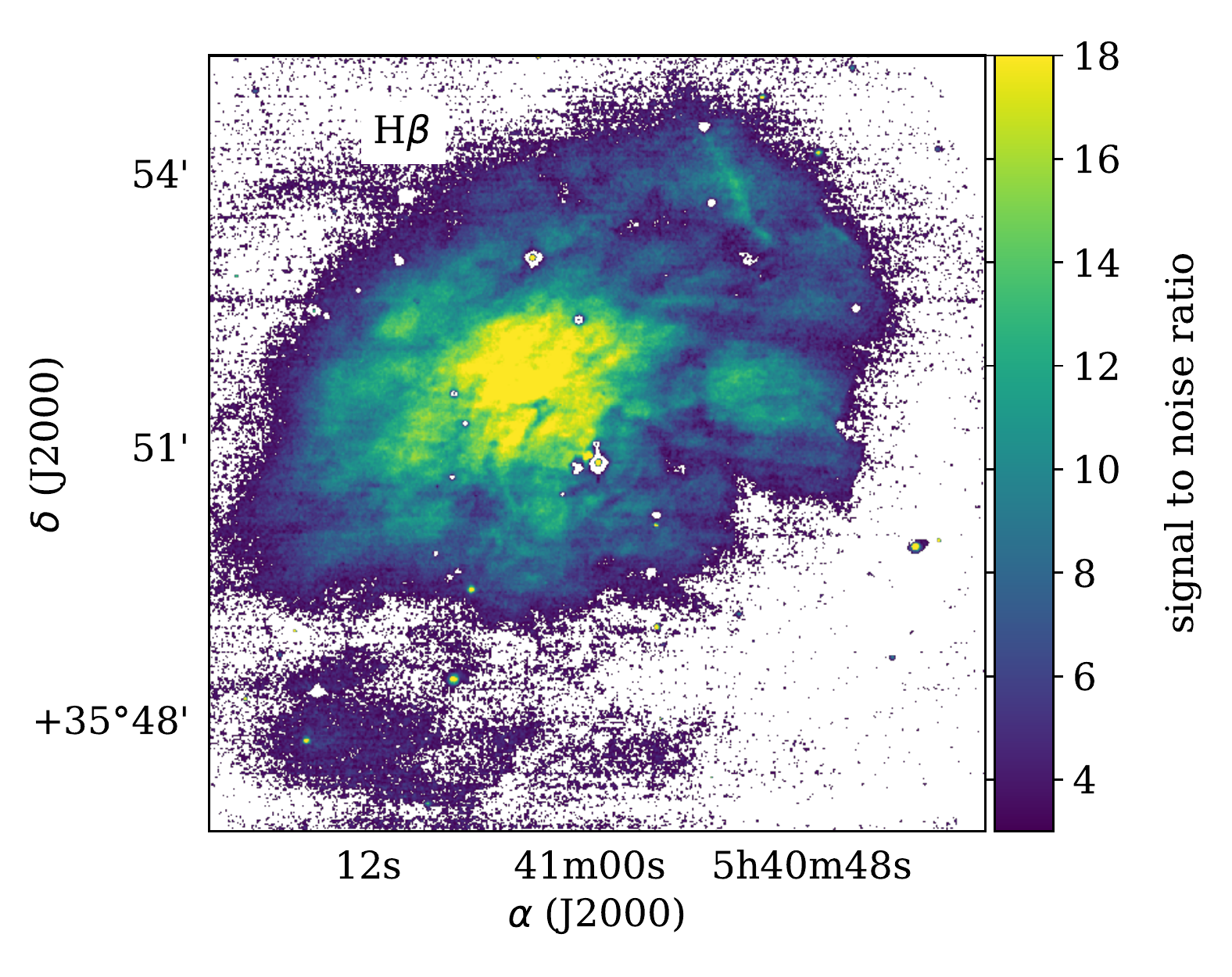}\\
	\includegraphics[width=\columnwidth]{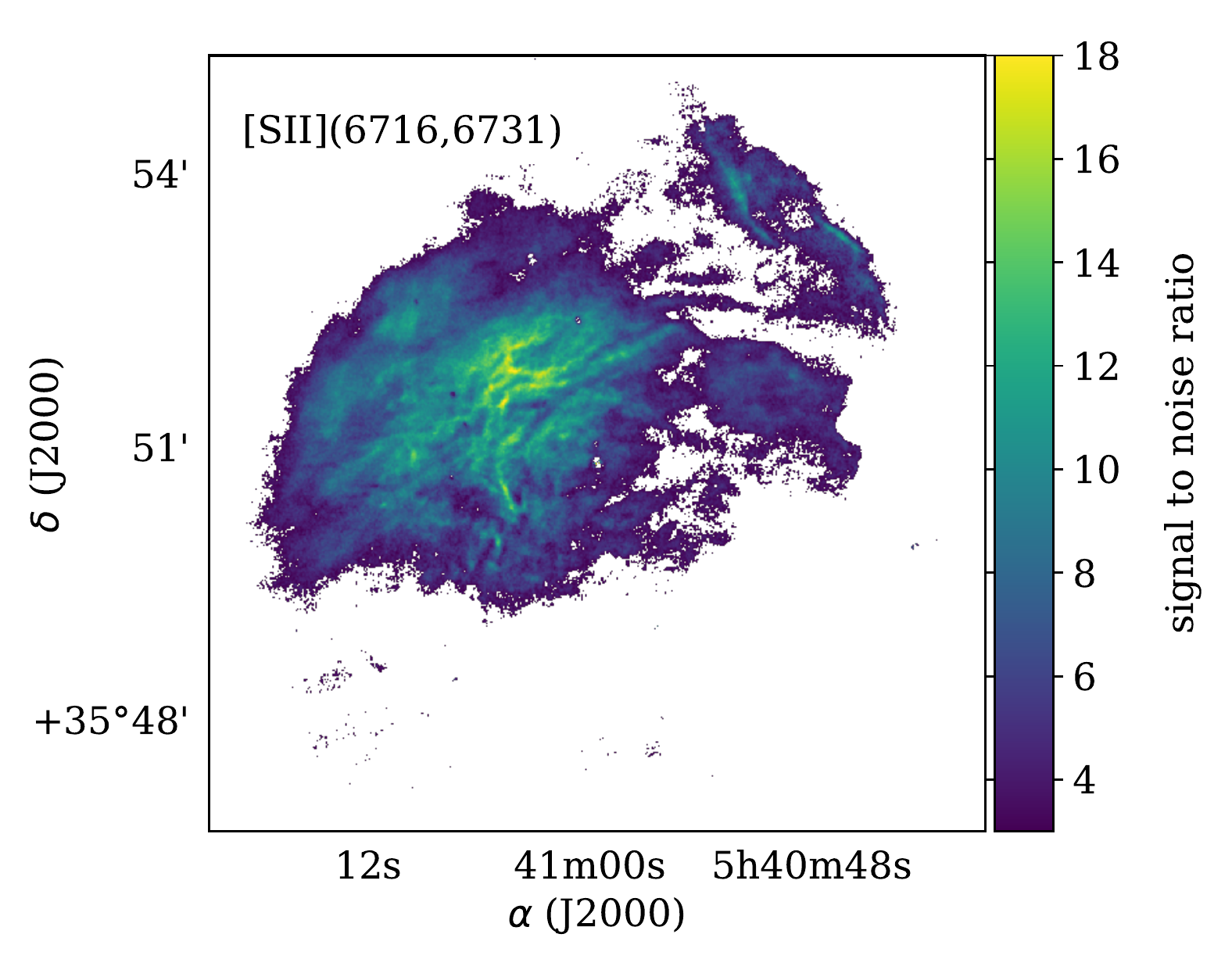}
	\includegraphics[width=\columnwidth]{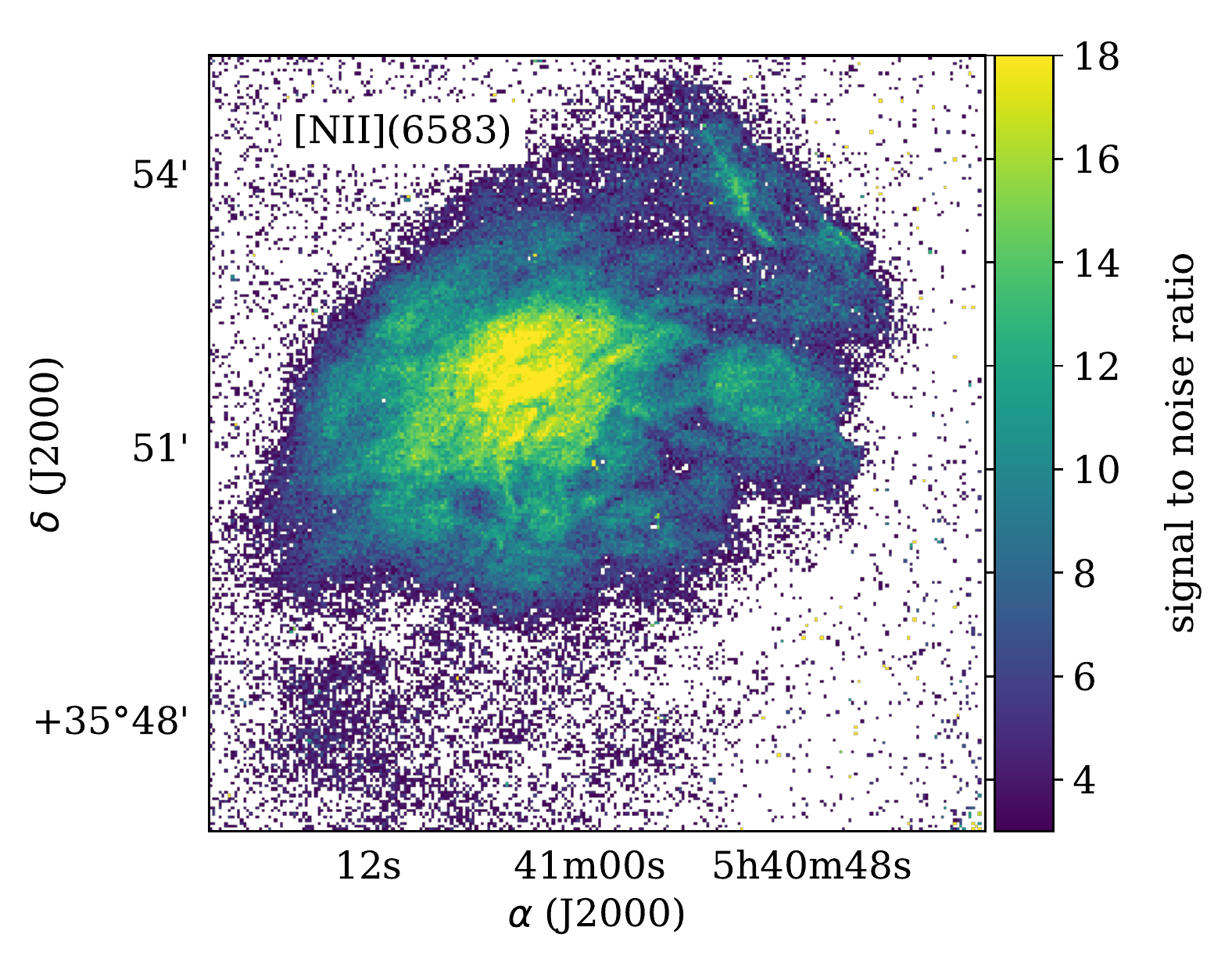}
	\includegraphics[width=\columnwidth]{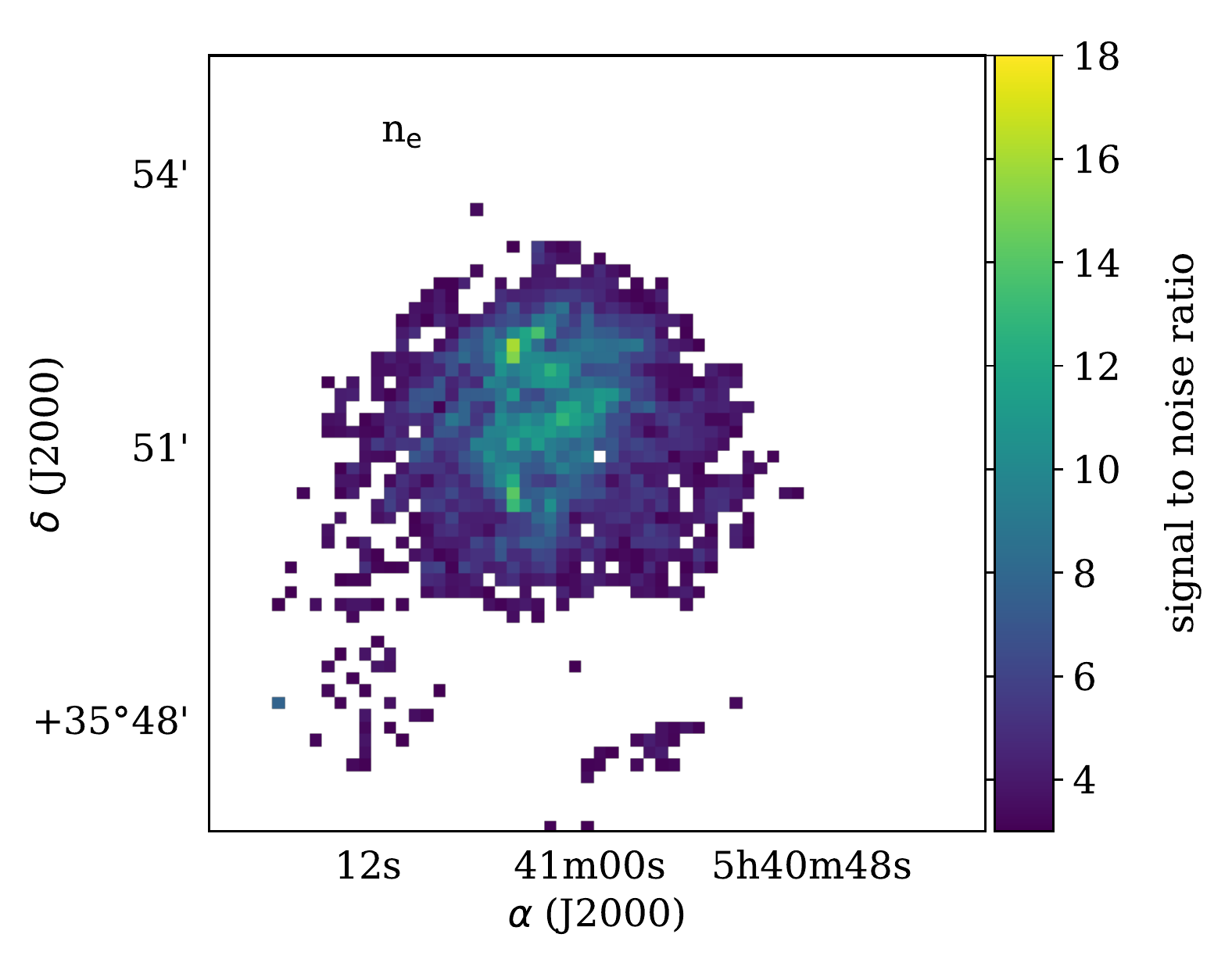}
	\includegraphics[width=\columnwidth]{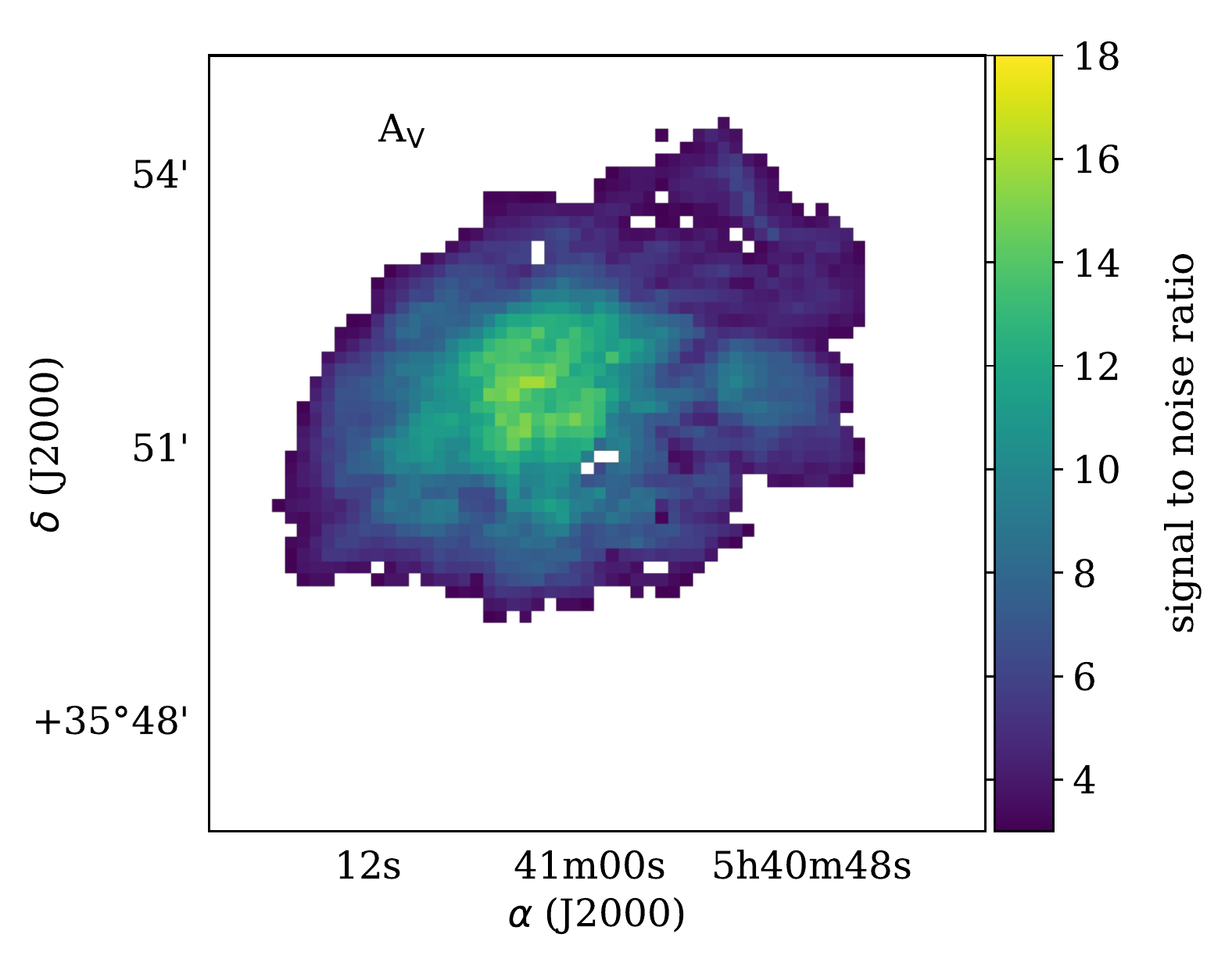}
	\caption{Maps of signal to noise ratio for pixels where the ratio is higher than 3. These pixels were used in the data analysis described in the paper.}
    \label{fig:err}
\end{figure*}

\bsp	
\label{lastpage}
\end{document}